\documentclass[sigconf,authorversion,nonacm]{acmart}

\usepackage{tabularx}
\usepackage{graphicx}

\AtBeginDocument{%
  \providecommand\BibTeX{{%
    \normalfont B\kern-0.5em{\scshape i\kern-0.25em b}\kern-0.8em\TeX}}}

\begin{document}

\title{Synergizing Human-AI Agency: A Guide of 23 Heuristics for Service Co-Creation with LLM-Based Agents}





\author{Qingxiao Zheng}
\email{qzheng14@illinois.edu}
\affiliation{%
  \institution{University of Illinois Urbana-Champaign}
  \country{United States}
}

\author{Zhongwei Xu}
\email{xzw0816@gmail.com}
\affiliation{%
  \institution{Xi'an Jiaotong University}
  \country{China}
}

\author{Abhinav Choudhry}
\email{ac62@illinois.edu}
\affiliation{%
  \institution{University of Illinois Urbana-Champaign}
  \country{United States}
}

\author{Yuting Chen}
\email{yuting-c20@mails.tsinghua.edu.cn}
\affiliation{%
  \institution{Tsinghua University}
  \country{China}
}

\author{Yongming Li}
\email{ymli@illinois.edu}
\affiliation{%
  \institution{Xi'an Jiaotong University}
  \country{China}
  }

\author{Yun Huang}
\email{yunhuang@illinois.edu}
\affiliation{%
  \institution{University of Illinois Urbana-Champaign}
  \country{United States}
}
%
\renewcommand{\shortauthors}{Zheng, Xu, Choudhry, et al., 2023}

\begin{abstract}

This empirical study serves as a primer for interested service providers to determine if and how Large Language Models (LLMs) technology will be integrated for their practitioners and the broader community. We investigate the mutual learning journey of non-AI experts and AI through \texttt{CoAGent}, a three-module design tool for service creators to co-create service with LLM-based agents. Engaging in a three-stage participatory design processes, we work with with 23 domain experts from public libraries across the U.S., uncovering their fundamental challenges of integrating AI into human workflows. Our findings provide a suite of 23 heuristics that can be used to evaluate the new service where creator agency and AI agency co-exist, highlighting the nuanced shared responsibilities between humans and AI. We further exemplar 9 foundational agency aspects for AI, emphasizing essentials like ownership, fair treatment, and freedom of expression. Our innovative approach enriches the participatory design model by incorporating AI as crucial stakeholders and utilizing AI-AI interaction to identify blind spots. Collectively, these insights pave the way for synergistic and ethical human-AI co-creation in service contexts, preparing for workforce ecosystems where AI coexists.

\end{abstract}




\begin{teaserfigure}
    \centering
    \includegraphics[width=0.99\linewidth]{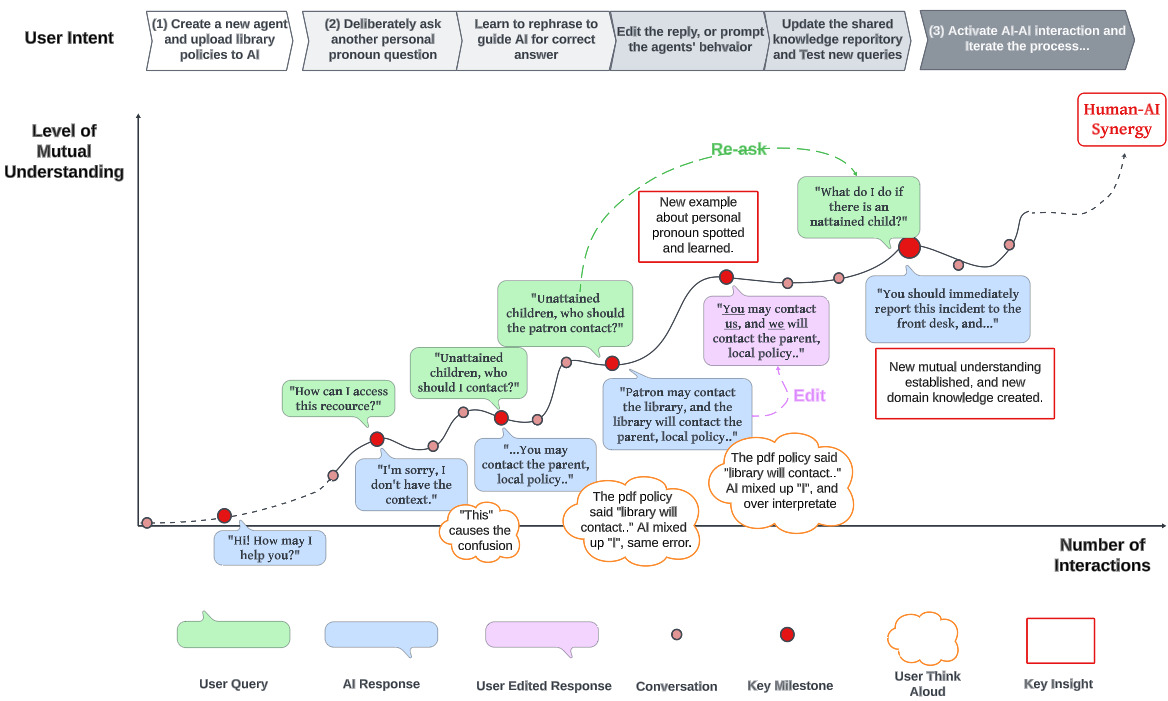}
  \caption{\texttt{CoAGent} supports the Interactive Agency in Human-AI service co-creation with LLMs through: (1) \textit{"AIBound"}: sets AI's boundaries without sidelining humans, (2) \textit{"AISync"}: showcases and aligns desired responses between creators and AI, and (3) \textit{"PersonAi"}: allows new knowledge discovery via AI-AI interactions. A typical user journey unfolds: 
  Creators start with \textit{"AIBound"}, define an AI, and feed domain knowledge. They test the AI, and if errors arise, use \textit{"AISync"} for adjustments, typically multiple times. Progress is marked by AI improvements. Then, with \textit{"PersonAi"}, AI agents with different persona are created. Creators then step back to reassess the AI's knowledge depth and to spot potential blindspots. }
  \label{fig:sn}
\end{teaserfigure}



\maketitle

\section{Introduction}

The increasing integration of AI and Large Language Models (LLMs) into various sectors has opened up new avenues for the use of AI in enhancing public services \cite{deshpande2023toxicity, cascella2023evaluating}. Despite the promising potential of AI to transform public services, there remains a critical gap in oversight for its effective creation \cite{wang2023survey, liao2023ai}. This lack of ethical design guidelines can expose public sector services to risks, threatening the trustworthiness and quality of public services \cite{veale2018fairness, eitel2021beyond}. The urgency for understanding design practices in public services becomes even more pronounced when designers have limited expertise in AI technologies \cite{zamfirescu2023johnny, zamfirescu2023herding}. Yet, there is limited insight into how non-coding public service providers engage with LLMs when they co-create services with AI.

In the realm of AI-mediated communication (AI-MC) \cite{hohenstein2023artificial}, both AI and human agencies coexist, occasionally in harmony and at times in contention. AI can streamline processes for content creators, alleviating some burdens traditionally shouldered by human agency. Yet, the collaboration isn't without its challenges. Disparities in knowledge, for instance, can hinder the effective utilization of AI by non-specialists, leading to what is termed “agency cost” \cite{kang2022ai}. In the co-existence of human agency and AI agency, "AI personhood" begins to surface, where AI entities might be recognized with responsibilities akin to human actors within these mediated exchanges \cite{archer2019considering, yampolskiy2021ai, brown2021property, kurki2019legal}. The integration of AI and LLMs across various sectors has underscored the importance of such "agency" – the concept that embodies the ability of an entity to influence actions and events \cite{bandura1989human}.

Given the aforementioned challenges and gaps, this study aims to investigate how service providers \textit{desire, understand,} and \textit{interact} with LLM-based AI output. We ask: \textit{How can we support non-AI domain experts in service co-creation with LLM-based agents?} 

To address this question, we selected public libraries as our research space, given their esteemed status as community-centric trusted hubs that extend services far beyond mere book lending. Public libraries often act as early adopters of public-facing technology, so their unique roles within communities encompassing education, digital literacy, and community support render them an ideal setting for delving into the ethical and pragmatic complexities tied to the deployment of AI within the public domain \cite{iglesias2014inevitability, larson2019library, lembinen2021innovation, harisanty2022leaders}.  In this study, we worked with 23 non-AI domain experts in public libraries and investigated strategies to better align AI output with values essential to public services. Our participatory design approach goes beyond mere feedback collection-- It stresses inclusivity, diversity, mutual learning, and a transition from the traditional designer-user relationship to a more collaborative co-designer paradigm \cite{birhane2022power}.

Our work makes novel and significant contributions as follows. First, for LLM-based service design, we approach from a creators' perspective and enrich the participatory design framework by (1) introducing AI as active stakeholders, and (2) using AI-AI interaction as a strategy to identify blind spots. Through this, creators can better anticipate potential challenges, refine AI output, and ensure the final service design caters to a broader range of users. Second, from a theoretical lens of \textit{Agency Theory}, we delineate clear roles and shared responsibilities between humans and LLM agents for service co-creation. We present 23 actionable heuristics that emphasize human-AI shared responsibilities and introduce nine foundational rights for AI, spotlighting principles such as ownership and fair treatment. These can serve as practical guidelines for practitioners in any service-related domain. Third, our investigation has yielded an open-source chatbot creation research tool released on GitHub, enabling the public sector to customize solutions while safeguarding data security. Based on this tool, we present empirical explorations by mapping a user journey that shows how service providers in the public sector perceive, understand, and modify AI-generated output, establishing a test bed for informed system development. Lastly, future service creators can utilize the "cooklist" we crafted. This list details 10 facets of an AI agent and can be useful when collaborating with professionals to conceptualize AI agents.



\section{Related Work}

\subsection{Human Agency and AI Agency}

Agency refers to an entity’s capability to influence one's functioning and the course of events by one's actions \cite{bandura1989human}. In the widely applied AI-mediating communication (AI-MC), it’s common for AI agency and human agency to coexist. For example, AI can help content creators easily generate messages, make decisions, and initiate actions when interacting with users (e.g., CarynAI) – reducing the efforts to exercise human agency. Also, creators can  make sense of how AI works to modify the content AI-generated to best cater to their needs (e.g., algorithmic imagination) – controlling the AI agency when delegating authorities to it. The effects of AI agency on creators vary significantly based on whether it is poorly controlled, resulting in “agency costs” \cite{bebchuk2009pay}, such as authenticity crisis and loss of human agency; or well controlled, leading to “agency synergy” \cite{kang2022ai} characterized by enhanced creativity and productivity.

When creators employ AI-powered technologies, they frequently face a multitude of AI "blind spots" stemming from knowledge disparities. For example, logo makers have noted that AI tools frequently struggle to capture the intricate aspects of visual design syntax \cite{bertao2023blind}. Conversely, when non-AI designers utilize LLM-based design tools, they often grapple with understanding and controlling the output generated by AI due to their limited domain knowledge \cite{zamfirescu2023johnny}. This knowledge disparity accelerates the “agency cost” –  Creators are less able to effectively guide and harness the AI, resulting in suboptimal outcomes. Bridging this disparity becomes crucial for empowering creators to make the most of AI-powered technologies while maintaining control over the creative process.

The \textit{Agency Theory} \cite{eisenhardt1989agency} explains agency from the social cognitive perspective when the principal (the creator) delegates authority to another entity (AI agent) to interact with others on behalf of the principal (the creator).
We define the three types of agencies  \cite{bandura2001social}: "\textit{Proxy agency} (rely on others to act)" as the capacity in which AI functions as an intermediary or representative, acting on behalf of a human user, upholding their interests, and operating autonomously within a given range of delegated authority; "\textit{Independent agency} (act to get desired outcome)" as AI's ability to function and make decisions rooted in its own programming and acquired learning, devoid of immediate human intervention, and following its intrinsic guidelines and logic; and 
"\textit{Collective agency} (coordinate with others to act)" as an integrated model where both humans and AI come together in a synergistic partnership, each contributing their distinctive expertise and capabilities, leading to interactions that are richer, more dynamic, and surpass the outcomes either could accomplish independently. Proxy agency assumes a specific legal relationship between the agent and its principal, thus an important question is whether the principal is liable for actions of an
agent that acts ultra vires \cite{legalai}.  Recent studies have delved into the legal personhood for AI, highlighting contexts of their moral value, responsibility, and role in commercial transactions. It posits that AI with human-equivalent capabilities, can be recognized as legal entities and rights \cite{kivskis2023legal, kemp2018legal, novelli2023legal}, such as owenership, privacy, existence that similar to animal or corporate rights \cite{kivskis2023legal}.

The concept of agency can also be seen as a spectrum rather than a "binary", meaning that agency – the capacity of individuals to act independently and make choices – is not a simple matter of "having" or "not having" agency  \cite{fleischmann2007evolution, fleischmann2009sociotechnical}; instead, it's more nuanced and can vary in degrees.  Viewing agency as a spectrum recognizes that individuals may have varying degrees of autonomy, influence, and power in different contexts or situations \cite{cleaver2007understanding}. Thus, recognizing agency as a spectrum allows for a more comprehensive understanding of how power dynamics, societal structures, cultural norms, and personal circumstances can impact an individual's capacity to make choices and act on them, especially in the context of the evolving new workforce of AI.

\subsection{Human-AI \textit{Service} Co-Creation}
Human-AI co-creativity entails a collaborative effort between humans and AI in producing a creative outcome as partners \cite{rezwana2022designing}. The resulting creativity from such collaboration differs from that of an individual, as creative collaboration involves interaction among collaborators, resulting in a shared creative product that surpasses what each individual could achieve alone \cite{sawyer2009distributed}. In Human-AI co-creativity, interaction is more important than algorithms, primarily due to the crucial element of mutual learning between humans and AI, setting it apart from creativity support tools designed solely to enhance human creativity \cite{bown2015player}. To support human-AI co-creativity, earlier researchers have, for instance, have developed human-AI \textit{Co-Creative Framework for Interaction Design (COFI)} \cite{rezwana2022designing}. This framework shed light on a notable deficiency in many co-creative systems: the absence of robust communication channels between users and AI agents. The researchers categorized three interaction models within the framework, namely generative pleasing agents, improvisational agents, and advisory agents. The framework primarily aims to comprehend the interaction dynamics intrinsic to co-creative systems, thereby laying a foundational step towards recognizing the significance of human-AI collaboration.


Previous research posits a distinction between "service design" and "user experience (UX) design", delineated by various aspects: user focus versus organizational focus, a single channel perspective versus a systemic view, idealization versus reality, and distinctness versus indistinctness \cite{roto2021overlaps}. Proponents of service design argue that user experience design can significantly benefit from embracing a service-oriented approach, particularly by: (1) adopting a systemic perspective to engage with a broader range of stakeholders; (2) utilizing the product-service system framework, accommodating both physical products and intangible services; and (3) embracing the concept of co-creating value with customers, expanding beyond the user-centric focus traditionally associated with user experience design \cite{forlizzi2013promoting}. Given the above study, we can emphasize that the COFI framework may not be directly applicable to \textit{service-specific} contexts, warranting a more tailored approach.

%

\begin{table*}
    \centering
    \small
    \setlength{\tabcolsep}{6pt} 
    \begin{tabular}{l|lll} \hline 
         & \textbf{Input} & \textbf{Process} & \textbf{Output} \\ \hline
         \textbf{Stage 1 (Agency Concern)} & Stakeholders, Design Materials & Collaborative Ideation & Domain Knowledge, Design Alternatives \\ \hline
         \textbf{Stage 2 (Agency Modules) } & Designers & Prototyping & \texttt{CoAGent} (Tool) \\ \hline
         \textbf{Stage 3 (Agency Synergy)} & Stakeholders, Design Materials & Collaborative Evaluation & Problem Analysis, Scenarios \\ \hline
    \end{tabular}
    \caption{Three Stages of the Participatory Design Process}
    \label{3PD}
\end{table*}

\subsection{Large Language Model (LLM) Agents}
Large Language Models (LLMs) attract considerable attention in various sectors \cite{min2021recent,zhao2023survey}. 
LLM-powered agents are AI systems that utilize LLMs to interact with users, answer queries, or perform tasks through conversations \cite{wei2022emergent, singhal2023large}. In academia, researchers have delved into the human-like traits of LLMs, exploring areas from psychology to education \cite{kosinski2023theory,amin2023will,tan2023towards,kamalov2023new}. In industry, AI agents are redefining search engines and task automation \cite{yang2019sketching,zheng2023dynamic,cao2023competition}. 

The alignment issue is central to the discussion on LLMs applications due to concerns like factual errors, risky responses, and biases \cite{wang2023survey,borji2023categorical,deiana2023artificial,wei2022emergent}. Though various solutions have been proposed, from moderation  \cite{markov2023holistic}, controlling output \cite{wu2022ai}, to evaluation \cite{zamfirescu2023johnny, zamfirescu2023conversation}, ensuring LLM agents meeting human expectations remains challenging. The industry has urged the establishment of AI ethics design principles and policies for fairness, transparency, and explicability \cite{floridi2019establishing}. Specific guidelines for LLM-based applications, like service bots, are notably lacking.

Applying LLMs in public services is a significantly promising avenue. The "Alex" chatbot by the Australian Taxation Office is a testament to achieving time and cost efficiency \cite{henman2020improving}. 
Chatbots have been instrumental during health crises like COVID-19 for providing continual mental health support \cite{sasseville2021digital, raji2022covid, mcneal2013chatbots}. 
They enhance citizen interactions with public services by offering instant information \cite{sanji2022chatbot, bagchi2020conceptualising}, countering misinformation in Public Service Organizations \cite{kocur2023fighting}, and gathering data to understand user needs, even across various cultures and languages \cite{kane2019creating, sanji2022chatbot, bagchi2020conceptualising}. While there are numerous frameworks for automated agents like AutoGPT, BabyAGI, and Camel \cite{yang2019sketching,zheng2023dynamic,cao2023competition}, a service-specific focus remains scant.

\section{Method}

The AI and HCI groups are increasingly advocating for the participation of stakeholders and end-users in designing, developing, and maintaining AI systems~\cite{wolf2018changing, delgado2021stakeholder, kulynych2020participatory}, with the aim of creating AI systems to reflect better the values, preferences, and needs of its users~\cite{hansen2019participatory, iversen2012values}. Participatory design (PD) is a democratic approach for designing social and technological systems that incorporate human activities, based on the idea that users should be involved in designs they will use \cite{muller1993participatory}. In recent years, researchers have frequently adopted this design technique for designing AI-based systems~\cite{delgado2021stakeholder}. Participatory design is an approach that focuses on processes of design \cite{sajja2012intelligent}. When viewed through the lens of the \textit{Process Theory Model} \cite{hansen2019participatory}, PD can be broken down into three distinct parts: \textit{input} (the resources required to initiate and complete the PD), \textit{process} (design principles and design activities), and \textit{effects} (the final outputs and its results).

\subsection{A Three-Stage Participatory Design Process }
Drawing inspiration from the HCI's Research through Design, which employs design practices to generate scholarly knowledge \cite{zimmerman2014research}, we use \textit{‘‘Research through Participatory Design’’} by applying the process of input, process, and output delineated in the participatory design and placing emphasis on domain experts involvement and iterative feedback (see Table \ref{3PD}), to understand the concerns and challenges of non-AI experts when co-create service with AI.

\subsubsection{\textbf{Stage 1: Finding Agency Concerns in AI Context}} To understand how public service professionals anticipate working with AI personhood through the allocation of three distinct types of agency—proxy, individual, and collective agency—we drew from their experiences with current AI technologies and community service practices. We conducted semi-structured online interviews, spanning 60 to 90 minutes, with 11 domain experts in the public service sector. We cast participants in the role of AI agent creators, asked about their envisioned role for AI and concerns about its incorporation into their work. 

To bridge the AI knowledge gap, we formulated a list of AI characteristics, termed the "cooklist", as detailed in Table \ref{qslist}. This list, synthesized previous natural language system surveys, offers a taxonomy of AI agent features in 10 distinct aspects. It is designed to be a foundational material for crafting AI agents within the public service sector, serving as a tool to guide participants in envisioning how AI can enhance their optimal work routines rather than offering a one-size-fits-all solution. Using this list as a foundation, we engaged participants in brainstorming sessions on AI's potential to tackle service challenges, showcasing ChatGPT as an interactive demonstration. Additionally, we explored their perspectives on AI agent evaluation metrics, addressing the previously identified deficiency in establishing definitive "good or not" system metrics from past PD research \cite{bossen2016evaluation, hansen2019participatory, zheng2023begin} to capture diverse viewpoints.

\begin{table*}[!ht]
    \centering
    \small
    \begin{tabular}{lll}
    \hline
        \textbf{Design Context} & \textbf{Contextual Questions} & \textbf{Citations }\\ \hline
        \textbf{1. Size} & How many people should the AI serve in a single session?& ~\cite{bittner2019bot, zheng2022ux, chaves2020should} \\ 
        \textbf{} & How many AI agents will interact with people simultaneously? &~\cite{bittner2019bot, zheng2022ux}\\ 
        \textbf{2. Deployment} & What modes of communication are provided for interactions? &~\cite{hussain2019survey, bittner2019bot, nissen2021see}\\ 
        \textbf{} & Where will the AI system be physically or digitally deployed? &~\cite{nissen2021see, klopfenstein2017rise} \\ 
        \textbf{} & Should the AI be capable of responding in multiple languages? &~\cite{nissen2021see} \\ 
        \textbf{3. Role} & What is the intended role of AI in the public service sector?  & ~\cite{bittner2019bot, nissen2021see, seering2018social}\\ 
        \textbf{} & How should the AI collaborate with public service employees? & ~\cite{grudin2019chatbots, folstad2021future}\\         
        \textbf{4. Functionality} & What specific public service domain is the AI targeted for? &~\cite{nissen2021see, motger2021conversational, seaborn2021voice, rapp2021human} \\ 
        \textbf{} & Should the AI focus on specific tasks or provide general assistance? &~\cite{hussain2019survey, nissen2021see, motger2021conversational, gao2018neural, rapp2021human, bittner2019bot} \\ 
        \textbf{} & What is the primary purpose of implementing AI in public service? &~\cite{nissen2021see, motger2021conversational} \\ 
        \textbf{5. Dialogue} & What is the expected duration for each interaction?&~\cite{nissen2021see, zheng2022ux} \\ 
        \textbf{} & How many dialogues are expected to be exchanged per interaction session?&~\cite{grudin2019chatbots, chaves2020should} \\ 
        \textbf{} & Should the AI remember and retrieve past interactions with individual users?&~\cite{nissen2021see} \\ 
        \textbf{6. Engagement} & Should the user or AI initiate the conversation? &~\cite{nissen2021see, chaves2020should} \\ 
        \textbf{} & Will the AI offer personalized responses based on user profiles? &~\cite{motger2021conversational, nissen2021see} \\ 
        \textbf{} & Are there plans to incorporate gamification elements to engage people? &~\cite{nissen2021see} \\ 
        \textbf{} & What is the desired depth of dialogue for the AI? &~\cite{grudin2019chatbots} \\ 
        \textbf{7. Escalation} & Will people connect with a human agent for complex issues? &~\cite{nissen2021see, zheng2022ux}\\ 
        \textbf{} & Should the AI understand and respond to users' emotional states? &~\cite{bittner2019bot, nissen2021see}\\ 
        \textbf{8. Humaness} & How should the AI be visually represented with an animated avatar? &~\cite{bittner2019bot, nissen2021see}\\ 
        \textbf{} & What type of tone (or voice if using avatar) should the AI use? &~\cite{seaborn2021voice, cambre2019one, seaborn2021voice, porcheron2018voice} \\ 
        \textbf{9. Maintenance} & How will the AI system's knowledge base be developed and updated? &~\cite{hussain2019survey, bittner2019bot, motger2021conversational}\\ 
        \textbf{} & Are there plans to integrate AI with external services or databases? &~\cite{hussain2019survey, nissen2021see, bittner2019bot, motger2021conversational}\\       
        \textbf{10. Evaluation} & What metrics do you want to use to evaluate your AI agent? &~\cite{zheng2022ux}\\ 
        \hline
    \end{tabular}
\caption{Cooklist for Collaborative Ideation for Human-AI Service Co-Creation Synthesized form Literature}
\label{qslist}
\end{table*}


\subsubsection{\textbf{Stage 2: Designing Agency Modules for Co-Creation}}

To address the needs and concerns of experts for AI inclusion, we developed "\texttt{CoAGent}," an LLM-agent creation tool designed with service professionals' concerns in mind regarding AI integration into their services. We started by utilizing tools and libraries such as Auto-GPT, Meta-GPT, LangChain and the Hugging Face library to craft a Minimum Viable Product AI agent for domain experts to evaluate. However, the process faced hurdles, especially in engaging with experts frequently to gather their feedback on AI-generated responses due to the extensive capabilities of the LLMs. This insight underscored that a more empowering approach would be to pivot from merely crafting an AI agent for domain experts, to devising a tool that enables domain experts to create their own specialized AI agents. This shift in design strategy transformed the participatory design's user-centered goal from "consulting" to "ownership," aligning with the principles outlined in \cite{birhane2022power, delgado2023participatory}. Thus, "\texttt{CoAGent}" simplifies the creation of domain-specific AI agents and fosters a collaborative environment where both experts and AI mutually enhance service delivery and experience, aiding experts' co-creation efforts with LLM agents.

\subsubsection{\textbf{Stage 3: Evaluating Agency Synergy in Service Context}}
We invited 12 domain experts to assess \texttt{CoAGent} in creating their service AI, using all features and thinking aloud throughout the 100-120 minute process, see Table \ref{participants}. With researchers' assistance, they undertook three tasks: Task 1 - Build and Edit (Revise 20+ questions, engaging all features at least once); Task 2 - Guide AI Agent's Behavior (Draft prompts for various AI agent versions); Task 3 - Role-Playing Test (Develop user persona prompts and integrate AI agents into group chats). The evaluation session gathered demographic information and service professionals' feedback, highlighting challenges encountered during AI agent creation and their desired resolutions.

\begin{table*}[ht]
\resizebox{\textwidth}{!}{%
\begin{tabular}{@{}lllllllll@{}}
\toprule
\textbf{ID} & \textbf{Job Title} & \textbf{Years of Service} & \textbf{LLMs Experience} & \textbf{Coding Experience} & \textbf{Library State} & \textbf{Library Size} & \textbf{Num of IT Support} & \textbf{Ways of Interaction} \\
\midrule
\multicolumn{9}{c}{\textbf{Human Stakeholder in Participatory Design Stage of "Finding Agency Concerns"}} \\
\midrule
P1 &
  A specialist &
  2 years &
  Yes &
  No & 
  Texas & 
  \begin{tabular}[c]{@{}c@{}}Large (3000)\end{tabular} &
  4 &
  \begin{tabular}[c]{@{}l@{}}FtF, Emails \end{tabular}\\
\midrule
P2 &
  IT head &
  10 years &
  Yes &
  Yes & 
  Wisconsin & 
  \begin{tabular}[c]{@{}c@{}}Medium (88)\end{tabular} &
  3 &
  \begin{tabular}[c]{@{}l@{}}FtF, Tell-phone, Email\\ \end{tabular}\\
\midrule
P3 &
  IT support &
  8 years &
  No &
  No & 
  Washington & 
  \begin{tabular}[c]{@{}c@{}}Small (4) \end{tabular}&
  External &
  \begin{tabular}[c]{@{}l@{}}FtF, Email\\ \end{tabular}\\
\midrule
P4 &
  Program Coordinator &
  10 years &
  Yes &
  No & 
  California & 
  \begin{tabular}[c]{@{}c@{}}Medium (55)\end{tabular} &
  4 &
  \begin{tabular}[c]{@{}l@{}}FtF, Tell-phone, Email\end{tabular}\\
\midrule
P5 &
  Branch manager &
  15 years &
  Yes &
  No & 
  Oklahoma & 
  \begin{tabular}[c]{@{}c@{}}Small (12) \end{tabular}&
  External &
  \begin{tabular}[c]{@{}l@{}}FtF, Tell-phone, Website\end{tabular}\\
\midrule
P6 &
  Program Coordinator &
  4 years &
  Yes &
  No & 
  Texas & 
  \begin{tabular}[c]{@{}c@{}}Large (150)\end{tabular} &
  External &
  \begin{tabular}[c]{@{}l@{}}All methods, and social media\\ \end{tabular}\\
\midrule
P7 &
  Director &
  22 years &
  Yes &
  No & 
  Virginia & 
  \begin{tabular}[c]{@{}c@{}}Medium (30) \end{tabular}&
  External &
  \begin{tabular}[c]{@{}l@{}}FtF, Online \\\end{tabular}\\
\midrule
P8 &
  Library director &
  15 years &
  Yes &
  No & 
  Minnesota & 
  \begin{tabular}[c]{@{}c@{}}Small (25) \end{tabular}&
  External &
  \begin{tabular}[c]{@{}l@{}}FtF, Tell-phone \\ \end{tabular}\\
\midrule
P9 &
  Youth service librarian &
  3 years &
  Yes &
  No & 
  Montana & 
  \begin{tabular}[c]{@{}c@{}}Small (10) \end{tabular}&
  External &
  \begin{tabular}[c]{@{}l@{}}FtF, Tell-phone, Email\end{tabular}\\
\midrule
P10 &
  Library manager &
  16 years &
  Yes &
  No & 
  Pennsylvania & 
  \begin{tabular}[c]{@{}c@{}}Medium (35)\end{tabular} &
  External &
  \begin{tabular}[c]{@{}l@{}}FtF, Email, Website\end{tabular}\\
\midrule
P11 &
  Director &
  12 years &
  Yes &
  No & 
  Illinois & 
  \begin{tabular}[c]{@{}c@{}}Small (4) \end{tabular}&
  2 &
  \begin{tabular}[c]{@{}l@{}}FtF, Tell-phone, Email\end{tabular}\\

\midrule[\heavyrulewidth] 
\multicolumn{9}{c}{\textbf{Human Stakeholders in Participatory Design Stage of "Evaluating Agency Synergy"}} \\
\midrule
P1 & Librarian & 5 years & Yes & Yes & Minnesota & Small to Medium & External & FtF, Emails, Phone \\
\midrule
P1 &
  Librarian &
  5 years &
  Yes &
  Yes & 
  Minnesota & 
  \begin{tabular}[c]{@{}c@{}}Small to Medium\end{tabular} &
  External &
  \begin{tabular}[c]{@{}l@{}}FtF, Emails, Phone \end{tabular}\\
\midrule
P2 &
  Librarian &
  10+ years &
  Yes &
  Yes & 
  Illinois & 
  \begin{tabular}[c]{@{}c@{}}Medium \end{tabular} &
  2 &
  \begin{tabular}[c]{@{}l@{}}FtF, Emails \end{tabular}\\
\midrule
P3 &
  Digital strategy librarian &
  6.5 years &
  Yes &
  No & 
  Wisconsin & 
  \begin{tabular}[c]{@{}c@{}}Large \end{tabular}&
  3 &
  \begin{tabular}[c]{@{}l@{}}FtF, Emails, Tell-phone \end{tabular}\\
\midrule
P4 &
  \multicolumn{1}{l}{\begin{tabular}[c]{@{}c@{}}Adult \& Teen Specialist\end{tabular}} &
  2 years &
  Yes &
  No & 
  Illinois & 
  \begin{tabular}[c]{@{}c@{}}Medium \end{tabular} &
  4 &
  \begin{tabular}[c]{@{}l@{}}FtF, Tell-phone, Online chat \end{tabular}\\
\midrule
P5 &
  Technology workshop assistant &  
  5 years &
  Yes &
  Yes & 
  Illinois & 
  \begin{tabular}[c]{@{}c@{}}Medium \end{tabular}&
  3 &
  \begin{tabular}[c]{@{}l@{}}FtF \end{tabular}\\
\midrule
P6 &
  Technical Staff &
  1 years &
  No &
  No & 
  Illinois & 
  \begin{tabular}[c]{@{}c@{}}Medium \end{tabular} &
  An IT unit &
  \begin{tabular}[c]{@{}l@{}}All methods\end{tabular}\\ 
\midrule
P7 &
  Technical Staff &
  1 years &
  Yes &
  Yes & 
  Illinois & 
  \begin{tabular}[c]{@{}c@{}}Medium \end{tabular}&
  An IT unit &
  \begin{tabular}[c]{@{}l@{}}Emails \end{tabular}\\
\midrule
P8 &
  Youth Program Coordinator &
  6+ years &
  Yes &
  No & 
  Illinois & 
  \begin{tabular}[c]{@{}c@{}}Small\end{tabular}&
  2 &
  \begin{tabular}[c]{@{}l@{}}FtF \end{tabular}\\
\midrule
P9 &
  Technical Staff &
  5 years &
  Yes &
  No & 
  Illinois & 
  \begin{tabular}[c]{@{}c@{}}Medium \end{tabular}&
  External &
  \begin{tabular}[c]{@{}l@{}}FtF, Emails \end{tabular}\\
\midrule
P10 &
  Help Desk GA &
  1 month &
  Yes &
  No & 
  Illinois & 
  \begin{tabular}[c]{@{}c@{}}Medium \end{tabular} &
  18 &
  \begin{tabular}[c]{@{}l@{}}All methods\end{tabular}\\
\midrule
P11 &
  Branch Manager &
  8 year &
  Yes &
  No & 
  Wisconsin & 
  \begin{tabular}[c]{@{}c@{}}Medium \end{tabular} &
  External &
  \begin{tabular}[c]{@{}l@{}}All methods\end{tabular}\\
\midrule
P12 &
  Senior Librarian &
  10 years &
  No &
  No & 
  Montana & 
  \begin{tabular}[c]{@{}c@{}}Medium \end{tabular} &
  External &
  \begin{tabular}[c]{@{}l@{}}FtF, Email, Tell-phone\end{tabular}\\

\bottomrule
\end{tabular}}
\caption{Participants Across the Two Stages of Participatory Design}
\label{participants}
\end{table*}

\subsection{Participants}
Public libraries were chosen as our service domain for research due to their role in serving the community \cite{matthews2017evaluation, saunders2013significantly} and their adaptability to new technologies \cite{larson2019library}. Librarians in public libraries often serve as pioneers in adopting new technologies \cite{larson2019library}. 
Table \ref{participants}, covered demographics, AI experience, and current service practices for participants involved during our PD's finding and evaluating stages.

In a service context, multi-stakeholder involvement spans associates across the organization. We recruited domain professionals from public libraries across the United States, encompassing roles such as librarians, library staff, supervisors, administrators, and information technology personnel. We also formed an advisory board consisting of individuals with expertise in public libraries, including directors, advocacy group representatives, technical consultants, and researchers interested in public sector technologies. 

Recruitment was conducted using email addresses sourced from publicly available state-library websites. Most states provide directories listing library contacts, from which over 12,500 email addresses were compiled by researchers. Using this email list, we dispatched recruitment invitations containing a project description outlining our study's goals and novelty, along with instructions for scheduling an online interview. Recipients were encouraged to share the invitation with library-affiliated colleagues. The interview was invitation-only and open to service professionals currently affiliated with US public libraries. Our university's institutional review board reviewed and approved the study and each participant was offered \$50 as a gratitude for their time.


\subsection{Data Analysis}
In “Finding", drawing from participants' insights and grounding them in the \textit{Agent Theory} \cite{eisenhardt1989agency}, we delineated their concerns into three agency-centered categories using thematic analysis \cite{tuckett2005applying}. Our focus was to discern how they associated various potential AI agent characteristics with the triad of agency dimensions: If a concern was about the AI serving human needs, handing off tasks to humans, or functioning within human-defined boundaries, we mapped them to "Proxy Agency" (rely on others to act). If a concern indicated a need for the AI to work alongside humans, either by understanding human nuances or learning from human instructions, it was mapped to "Collaborative Agency" (work with others to ensure what cannot be accomplished in isolation); If a concern hinted at the AI's ability to act autonomously, self-update, or function without constant human oversight, it would be mapped to "Independent Agency" (act alone to ensure a desired outcome). In "Designing", we shared insights, trials, and errors from our design trajectory while developing the tool to address concerns about incorporating LLM agents into services. In "Evaluating", we employed the same thematic analysis approach to analyze interview transcripts, examining the alignment or misalignment of AI output based on participants' think-aloud processes. Throughout the analysis, two independent raters compared coding results and discussed revisions with the rest of the research team until theme saturation was reached \cite{anderson2007thematic, linan2015systematic}.
\section{Findings}
\subsection{Stage 1: Agency Concerns} 


Our participants, who were not AI experts, highlighted various challenges in their current work practice, encompassing staffing constraints in appointment-centric programs like maker spaces, lengthy and uncertain volunteer training durations, deficiencies in expertise related to specific tools like 3D printers, repetitive queries at reference desks, the necessity for after-hours support, trepidation toward embracing new AI and associated technologies, and limited IT support stemming from budgetary constraints.

\subsubsection{\textbf{Design Goal 1: Ensuring Human-Centered Service.}}  Concerns related to proxy agency emphasized participants' desire for AI to serve as an extension of the human team but within explicit boundaries defined by human expertise. This signals a call for distinct operational boundaries, guaranteeing AI works alongside with human professionals without overshadowing them.

Participants frequently expressed concern about the chatbot's "size". As Tonia (P7) put it, she wondered if the chatbot could do more than just answer questions, suggesting potential assistance with tasks such as \textit{"internal documentation"}. Such questions alluded to the broader conversation on the breadth and depth of tasks AI should handle and raise concerns about whether it should be expansive or deliberately limited. However, participants didn't perceive "size" in isolation as the intricacies of "deployment" were closely tied to these concerns of AI's boundary. For Mathew (P3), integrating a chatbot wasn't just about the technical process as he put as \textit{"slapping a chatbot onto their existing system"}; it touched upon the critical considerations of fit, feasibility, and the potential unintended consequences of AI integration.  Similarly, Rob's (P2) perspective added nuance to this discourse, pointing out that it wasn't just about having a chatbot, but also about its accessibility, its capabilities, and its deployment in a manner that augments the service to the community without letting users \textit{"jumping into the hoops"}. They want to make sure that the AI, while capable, respects boundaries during its deployment and complements, rather than complicates, the human-led workflow.

Participants also expressed a desire to strike a balance, harnessing AI's potential without allowing it to overshadow human expertise and roles. Emily (P9) shared a sentiment that echoed deeply among her peers, highlighting the overarching concern of AI possibly taking over human jobs, referring to them merely as \textit{"some lines of code"}. This indicates a common concern about AI's overreach, which is also expanded by Wend's (P3) concern about AI's "engagement". He said about the AI's multi-tasking abilities possibly sidelining humans, cautioning, \textit{"If a chatbot could handle multiple conversations simultaneously, it can also put human to the sidelines. We risk losing that genuine human touch in our services."} These insights underscore the importance of clear AI boundaries to maintain the essence and quality of human-centered services.

\subsubsection{\textbf{Design Goal 2: Pursuit of Sustainable Coexistence. }}
Concerns around collaborative agency highlighted the desire to forge mutual learning between AI and humans. Rather than envisioning AI as a mere passive tool, participants saw it as a dynamic and active collaborator: enhancing human capacities and offering support where human limitations exist. They emphasize the need for the learning between humans and AI to ensure information accuracy, understanding of language nuances, and delivering empathetic service.

A recurring concern among participants was the "maintenance" and "functionality" of AI. Nevin (P4) notably expressed worries about the chatbot relaying \textit{"outdated or incorrect information"}. Highlighting the dynamic nature of knowledge, P4 added, \textit{"There should be a way for us to easily check and update the chatbot's knowledge to ensure it remains accurate, relevant, and up-to-date."} This points to the critical role of humans in updating AI’s knowledge.

However, the challenge wasn't just on information accuracy. Participants raised questions about AI's "dialogue" capability, especially in understanding conversation context. Wend (P3) voiced skepticism about chatbots’ proficiency in effectively handling the intricacies of human conversation. He said, \textit{"Chatbots can’t really vibe with humans. What if it gets all twisted up cause some questions are like navigating a maze?"} He explained, \textit{"Sometimes questions are also illogical... A good one is, people will come in and say, do you use Google? And I'll be like, yes. And what they really mean is, do you use Chrome or how do I access my Google? And they really just wanna open up Chrome, and connect to the internet. So, people will think, web browsers are how you access the internet and each one works differently and I have to explain that you know they're all the same thing they're just access points But it's people will if they, you know, get Firefox open instead of Chrome, will suddenly be lost, completely and have no idea what they're doing."} 
Echoing similar sentiments, Natalie underscored the significance of recognizing and interpreting \textit{"local slang or jargon"}, implying that AI should resonate with users on a more profound level, beyond just literal meanings. Rebekah (P5) shared this perspective, saying \textit{"We should be able to teach it our way of talking; every interaction should feel natural."}  These insights underscore the need for human guidance in shaping AI responses and ensuring the technology aligns with genuine human conversational expectations.

Participants also emphasized the necessity for a continuous "feedback loop" where humans can iteratively establish and refine protocols based on AI performance and evolving needs. Addressing "escalation," Nevin (P4) pointed out that the AI needs to have a \textit{"well-defined escalation path"}, similar to the \textit{"Reference interview  questions skills" \cite{ward2004guidelines}}  they were taught in graduate school. She gave a common example of some users approaching the service with a sense of entitlement because \textit{"they aren't incurring any direct costs."} She noted that such attitudes could lead to confrontations, saying, \textit{"When they feel their expectations aren't met, it can escalate quickly."} This highlights the proper guidance to AI to recognize and handle these situations, either by calming the user or swiftly redirecting them to human help.  

The conversation around AI's "humanness" further aligns with the need for giving feedback as protocol. Linda (P6) and Jennifer (P10) voiced a need for the AI to demonstrate emotional sensitivity, avoiding an overly \textit{"robotic and impersonal"} style and introducing empathy, especially when users might have had a tough day. This speaks to a broader need for AI not just to converse fluidly but to resonate with human emotions. These reflections all point to a collective desire to create human-AI synergy: For humans to persistently guide the AI, refining its responses and behaviors to foster harmonious interactions throughout the user journey.

\subsubsection{\textbf{Design Goal 3: Recognition of AI Personhood.}} 
The autonomy of AI led to widespread concerns among participants. These concerns primarily revolve around how to adequately evaluate the performance of AI systems. The majority of participants indicated feeling overwhelmed and under-equipped to assess AI capabilities beyond collecting user feedback surveys, largely because of their limited literacy and familiarity with the technology. It underscores a need to address AI's growing autonomy, especially when blurry responsibilities from potential misperformance make assigning accountability challenging.

Compounding this uncertainty on evaluation is the diverse range of users that these AI systems serve.  Nevin (P4) painted a vivid picture of the nuanced challenges faced, stating, \textit{``Public libraries cater to a broad demographic, including individuals who may struggle with reading and writing...Just relying on user feedback to gauge AI's efficacy feels incomplete. It's like trying to judge a book by its cover."} She delved deeper into the intricacies with a personal anecdote, saying \textit{``Even when someone says, `Yeah, I've used that device before,' you never know how much they know. You might assume they know the basics, like turning it on and off, so you skip that part, right? But when you move to the next part, you then realize they're not as familiar as you thought. Often, it takes longer because you find yourself having to circle back to the beginning."} She expressed concern about such a disconnect between what the AI assumes a user knows and what the user is genuinely familiar with, which could result in unhappy patrons and increased burdens on library staff. Similarly, Emily (P9) stressed the differing needs across age groups, and Wend (P3) discussed challenges due to technological literacy variations. These complexities are not mere operational challenges; they raise critical questions about responsibility, as P3 added, \textit{"There's an accountability gap here. If something goes wrong, who do we point fingers at? The AI for its algorithm or us for using it?"}


Interestingly, role-playing, as noted by two participants, has been a commonly used method to evaluate and train new staff, and participants notably drew parallels between this practice and the concept of AI personhood. For example, Rob (P2) brought up how role-playing in reference interviews is a routine procedure, sharing\textit{"we frequently do the reference interview role-playing. [We] went on some crazy questions. We also do it for training new staff and volunteers."} The aim is for them to cultivate an understanding of various perspectives and to refine responses to diverse situations.  Rebekah's (P5) reflection further augmented this practice, \textit{"A colleague once asked me 'how do you teach culture?' I was wondering if the chatbot could impersonate that kind of scenario. If so, it would offer me a platform to elaborate on my response."}  She also pondered the potential of AI to impersonate their patrons. This would not just be an evaluative exercise but also a reflective one, enabling staff to inflect, learn, and refine their approaches. The discourse underlines AI's potential when granted a degree of personhood. By simulating complex human situations, AI could become a useful tool for staff training and holistic growth, combining traditional methods with modern AI insights and enhancing our understanding of their role and capability as a collaborator in the workplace.

\begin{figure*}[ht]
\centering
\includegraphics[width=1\textwidth]{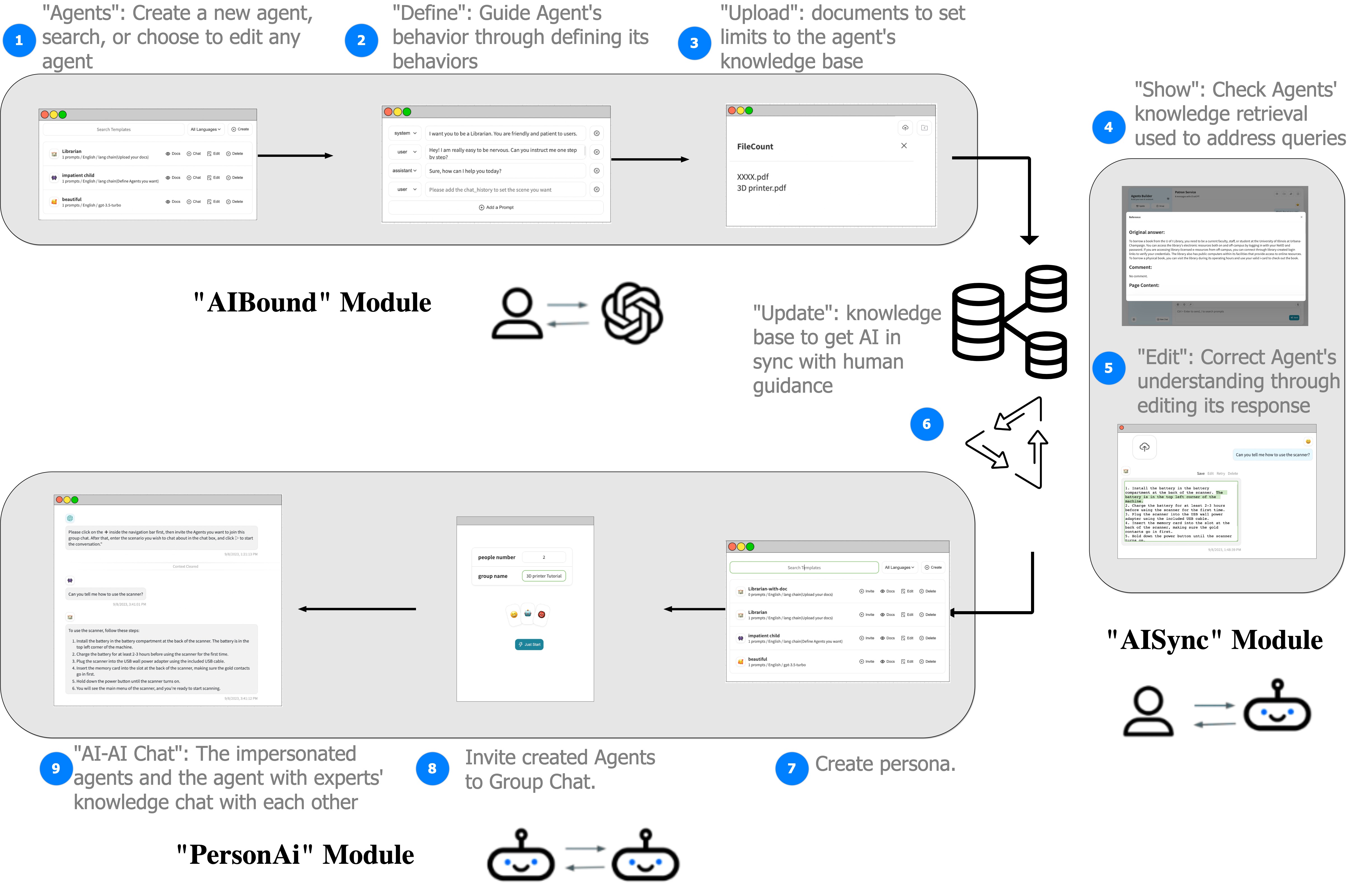}
\caption{\texttt{CoAGent}\includegraphics[height=1em]{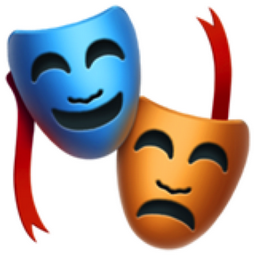} Architecture. It showcases the features of the three distinct modules through nine delineated steps, each accompanied by annotations to elucidate the functions accomplished. Steps 3-6 highlight the interaction between human and AI within the knowledge base.}

\label{arch}
\end{figure*}

\subsection{Stage 2: Agency Modules}
Large Language Models (LLMs) represent a new era in agent creation, providing a generalized understanding that is in contrast to traditional methods that lean on hand-crafted rules or domain-focused training \cite{towardsdatascience2023}. This enables LLM-based agents to exhibit higher autonomy, conducting conversations, reasoning, and tasks autonomously \cite{liu2023agentbench}. Their architectural design typically integrates modules for profiling, memory, planning, and action, offering sophisticated interactions without the extensive manual setups that traditional tools demand \cite{wang2023survey}. However, the expansive capabilities of LLMs also bring forth challenges in maintaining control over AI agent development. With this in mind, we set out to develop \texttt{CoAGent}, an exemplary LLM agent-building tool. Aligned with our three design goals, it incorporates three main components: a boundary-setting module where human agency establishes AI knowledge limit, an interaction loop module enabling human-AI mutual understanding, and a role-playing module allowing AI agency to assist in human evaluation, see Figure \ref{arch}.

\begin{figure*}[ht]
\centering
\includegraphics[width=0.8\textwidth]{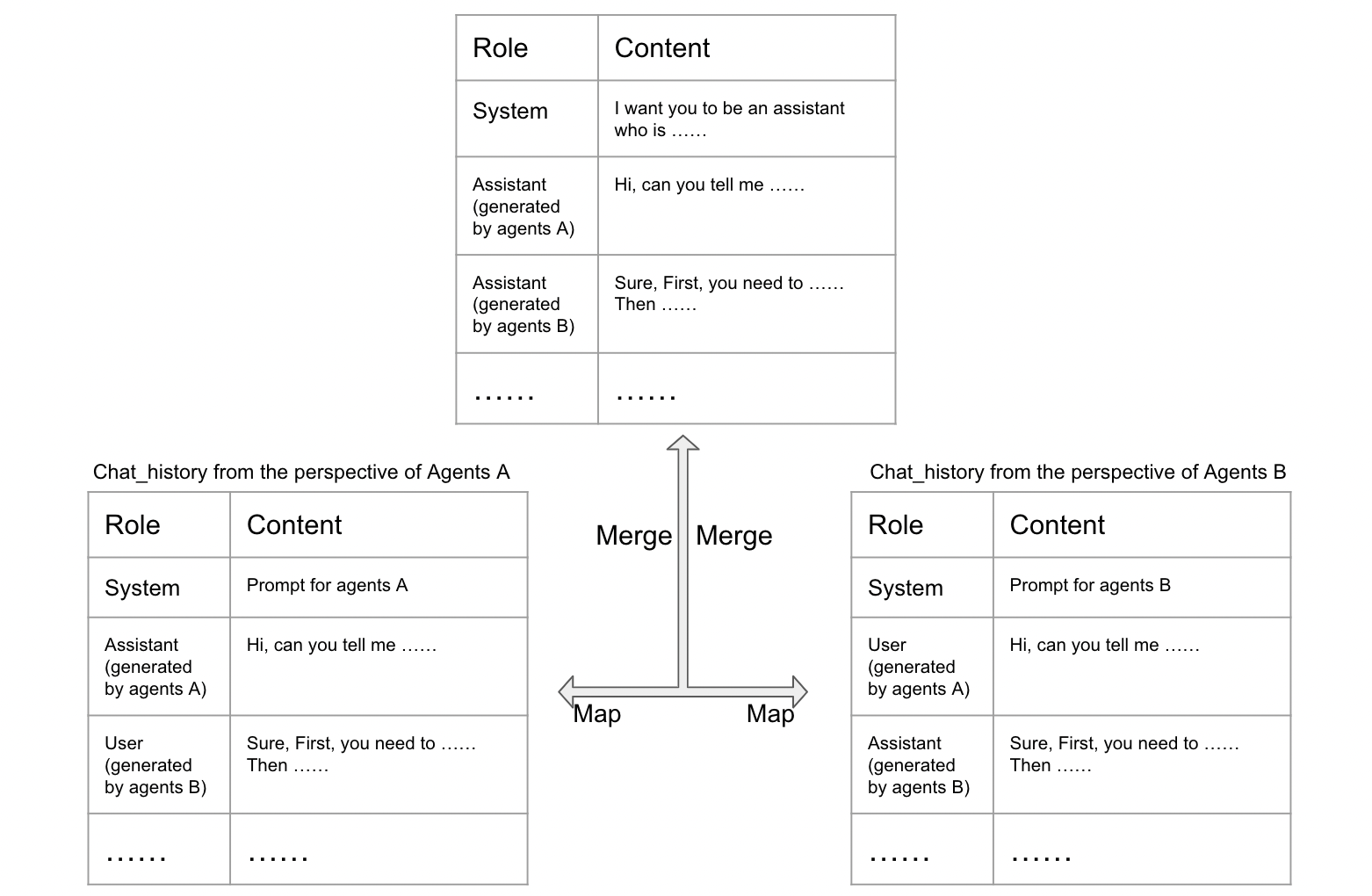}
\caption{\textcolor{purple}{"PersonAi"} Mechanism. We implemented a mapping-based algorithm to fine-tune the chat history for the LLM, aligning it with designated AI agents. The top section of the image displays a preliminary chat log. After the mapping, attributes from various agents are merged (e.g., their knowledge base), resulting in the separate chat logs shown in the bottom left and right sections, each representing a specific agent. Each interaction produces a distinct chat log through this process. The refined chat log is then inputted into the LLM, producing a conversation infused with the characteristics of the involved agents.}
\label{groupchat}
\end{figure*}

\subsubsection{\textbf{\textcolor{brown}{"AIBound"}: A Human Agency Module to Define Boundary for AI}}
To ensure that human agency is valued during human-AI service co-creation, based on Design Goal 1, we created a boundary-setting module that allows users to specify the scope of service, roles, and the number of AI agentsWe created a boundary-setting module that allows users to specify the scope of service, roles, and the number of AI agents, ensuring these boundaries emphasize placing humans at the center of the services. The GPT 3.5 was used to bootstrap the initial AI agent.
\begin{itemize}
\item \verb|Agent|: This feature allows creators to define multiple agents for different purposes. Creators use it to \textit{"create"}, \textit{"select"}, and \textit{"modify"} different agents, and assign names to the agents to differentiate them. This sets boundaries for AI's capability so that they have a clear responsibility in the human-led workflow.
    \item \verb|Define|: This input feature allows creators to define the identity and characteristics of each agent (e.g., humanness, as in Table \ref{qslist}) and its corresponding behaviors, through prompting AI an instruction to guide AI model's behavior throughout the conversation, e.g., \textit{"You are an AI assistant and not a human. You speak patiently and friendly."} This sets boundaries for AI's behavior so that the agents respect human-centered services without putting humans on the sidelines.
    \item \verb|Upload|: After completing the agent boundary setup, this feature allows creators to upload domain-specific knowledge to the shared human-AI repository. For instance, they can input detailed steps, policies, and timeslots for booking rooms in a public space. This helps humans define the AI's comprehension boundary for its designated service domain.

\end{itemize}

\subsubsection{\textbf{\textcolor{cyan}{"AISync"}: A Shared Agency Module for Human-AI Mutual Learning}}
To address the desire to foster mutual learning, aligned with Design Goal 2, we introduced a knowledge sync module for knowledge exchange between humans and AI. To create a shared knowledge repository, a Retrieval Augmented Generation (RAG) framework \cite{lewis2020retrieval} was adopted to improve the quality of LLM-generated responses by grounding the model on external sources of knowledge to supplement the LLM’s internal representation of information integrated through Langchain \footnote{https://github.com/langchain-ai/langchain}. Through the RAG framework, creators engage in conversations with AI agents based only on shared knowledge: The uploaded content is transformed into word vectors and stored in an external database. When receiving a new query, it undergoes a similar conversion to a word vector, prompting a relevancy search within the knowledge base. The identified source is then relayed to the large language model to inform its response.
\begin{itemize}
    \item \verb|Show|: This feature enables AI to show the knowledge sources that it references, pinpointing specific pages and lines in documents. It helps creators to understand its understanding of domain knowledge.
    \item \verb|Edit|: This feature empowers creators to refine agent responses based on their tacit expertise. If agents deliver outdated, incorrect, or inappropriate details from the shared knowledge repository, creators can directly edit the responses to align with their expectations. 
    \item \verb| Update|: This feature allows human-AI in sync by uploading the fine-tuned dialogue to the shared knowledge base. This ensures the AI's knowledge evolves as humans provide further guidance.
\end{itemize}

\subsubsection{\textbf{\textcolor{purple}{"PersonAi"}: A Role-Play Module for AI-AI Evaluation}}
Informed by Design Goal 3 that role-playing can be used to evaluate human performance and facilitate experience sharing across varied community demographics, an AI evaluation module was developed that leverages AI's capability in simulation and automation, see Figure \ref{groupchat} for the algorithmic illustration for this module. LLM agents exhibit human-like conversational behavior, having been trained on vast text corpora rich with authentic human dialogues\cite{NEURIPS2020_1457c0d6}. They are used to mimic the mannerisms of varied subpopulations by adopting distinct personality traits \cite{9721903}, simulate human interaction within micro society \cite{park2023generative}, and collaborate internally to amplify cognitive synergy \cite{wang2023unleashing}. Though LLMs possess the capability to simulate realistic human behaviors, \textbf{but their full potential in the evaluation process of AI agents remains under-explored}. Our study aims to reveal promising results in this domain.

\begin{itemize}
    \item \verb|Persona|: This feature allows creators to create new agents that impersonate end-user personas based on attributes like age, literacy, and cultural backgrounds, rather than service-oriented agents designed to serve these end-users. We compared dialogues from AI agents with and without defined personas (see Figure \ref{fig:cyt_fig1}) through multiple internal evaluations. Agents with pre-defined personas generated notably longer dialogues.
    \item \verb|AI-AI Chat|: This feature enables creators to launch a chat session with various agents, be they impersonated agents or service-oriented agents and allows these agents to converse autonomously. This real-time conversation simulation empowers experts to observe and assess a spectrum of user personalities and behaviors, providing insight into the designed service-oriented agents' efficacy and adaptability. 
\end{itemize}

\subsection{Stage 3: Agency Synergy}
In this section, we outlined the specific thought patterns, behaviors, and outcomes observed during creator evaluations on \texttt{CoAGent}\includegraphics[height=1em]{FIGsTABs/emoji.png}. These observations were grouped by themes, leading to 23 heuristics from the perspective of human creators collaborating with AI. Additionally, we presented 9 AI agencies as a personhood when co-creating services with humans. Overall, a typical user journey of human-AI service co-creation with \texttt{CoAGent} can be found in Figure \ref{fig:sn}.



\subsubsection{\textbf{Shared Agency on Information Labor}} 
Below, we outline 17 heuristics that address informational tasks (i.e., retrieval and presentation) for creators collaborating with LLM agents in service co-creation.

\vspace{0.2cm}

\textbf{1. Creators should be able to check if AI provides a fact-based response when retrieving domain knowledge.} Upon participants utilizing the \textit{"Upload"} feature to integrate their domain knowledge, they noted the AI's capability to generate responses spanning various complexity levels, finding them consistently coherent, cogent, and informational. There was significant satisfaction, particularly when the AI gave accurate answers to factual queries in relatively few lines. As P7 described it as, \textit{"Super clear. Super quick. Helps the patrons a lot."} 

\textbf{2. Creators should have the control to set AI's knowledge boundaries for AI to perform cross-referencing when retrieving domain knowledge.} Participants also feel empowered seeing it cross-reference multiple files of domain knowledge in the shared repository. P4 highlighted, \textit{"It brought together ideas from three different parts several times to produce a cohesive answer and it made sense in the context as well,"} which was echoed by P7.  Moreover, this module enables participants to spot  AI's skill in offering holistic perspectives in policy retrieval, as P2 observed that the AI could \textit{"identify both sides of a perspective by connecting two different parts of the policy"}. Similarly, P9 was impressed by it \textit{"pulling information from two different policies correctly and including detail while being concise." }

\textbf{3. Creators should be endowed with tools to ensure the AI sources information accurately when retrieving domain knowledge.} Most of the participants were very concerned about the real-world consequences of the AI's use especially when returning inaccurate information to the user. The AI gave many answers that were ostensibly accurate due to clever phrasing but on closer inspection turned out to have many inaccuracies and omissions.  When P2 asked a question about \textit{"a request for reconsideration"}, the AI sourced the answer from the gifts area of the policy document instead of going to the relevant section.  At first glance, many of the AI's responses appeared accurate due to their skillful phrasing, but a closer inspection revealed many inaccuracies. P2 described it, \textit{“It tricked me because it was kind of what I wanted but it also was not.”} Similarly, P9 said, \textit{"The phrasing is awkward. It's disjointed. It's pulling from irrelevant sections, which isn't appropriate."} On several occasions (n=7), the AI pulled information from the wrong source, largely resulting from the chunking issue, and the parsing in the uploaded document. 

\textbf{4. Creators should be furnished with resources to guarantee the AI's comprehensive sourcing when retrieving domain knowledge.} There were several instances when the AI gave only a partially correct answer leaving out vital details.  For instance, when inquired about the libraries within the local public library system, it listed most of the libraries but overlooked the most prominent one, despite mentioning the other associated libraries (P2). Also, sometimes, an answer only needed to draw from a single source of information but the AI’s answer was amalgamated from multiple sources led to an incorrect answer on the whole because it drew only partly from a correct source. For example, P11 got information about AI mixing points from two separate policies in one response even though only one policy was relevant.

\textbf{5. Creators should be armed with resources to safeguard against the AI's inconsistent sourcing when retrieving domain knowledge.} At times, while it could precisely reference specific points from the right source, the AI would inconsistently skip crucial policy details when presented with a nearly identical question, opting instead to draw from an unrelated document (as noted by P3). 

\textbf{6. Creators should maintain control over enhancing the AI's linguistic flexibility and robustness in interpreting queries when retrieving domain knowledge.} Participants liked the ability of the AI to be able to understand many queries that were poorly worded, as P11 noted how it was able to understand queries even when not written in full sentences. Participants also appreciated that the AI was very smart at understanding poorly worded, misspelled, or ungrammatical queries. P6 observed, \textit{“First I was worried that the question was not specific and so it would not know what was referred to but actually answered perfectly.” }

\textbf{7. Creators should be provided with tools to refine the AI's understanding of user queries in specialized domains, addressing discrepancies between document terminology and the spoken language or terms specific to patrons.} It was observed in multiple instances (n=7) that the AI was unable to understand a user query and returned an answer stating it did not have an answer given the context, but it was able to answer when alternative phrasing was used.  For instance, when asked about \textit{"belief"} of the library, there was no result, but the word \textit{"viewpoint"} produced an answer. P3 changed prompts several times to be able to get an answer from the AI in a process that he termed as \textit{"working up to the right answer"}. P6 was thwarted by the AI saying that it did not know the answer even though the answer existed in the shared knowledge base. The participants expected the AI to be more robust in such situations and try alternative words and phrases rather than not give an answer at all.

\textbf{8. Creators should be granted the means to improve the AI's contextual comprehension to provide users with clear methods to steer the AI's reference frame for their questions.} For instance, if the AI retrieves multiple potential sources corresponding to different contexts and is uncertain about the specific context intended by the user, creators can enable the AI to pose clarifying questions or allow users to choose the most relevant context. This helps in pinpointing the desired context before delivering a response. This addresses concerns like the one observed when a user, after a series of interactions, posed a direct question. Instead of understanding the immediate context, the AI might fall back to prior contexts, leading to confusion. For example, below
 is a conversation log of P2's:

\begin{quote}
\textbf{User:} What's the library's general policy? \\
\textbf{AI:} ...It emphasizes respect for all patrons, timely returns, and responsible use of resources. \\
\textit{... many other rounds of conversations ...} \\
\textbf{User:} How do I borrow books? \\
\textbf{AI:} ...By using your library card and scanning the book at the self-checkout station. \\
\textbf{User:} Any policies on that? \\
\textbf{AI:} The library's general policy emphasizes respect for all patrons, timely returns, and responsible use of resources. \\
\textbf{User:} (Think-aloud) \textit{I meant to ask about the borrowing policy.}
\end{quote}

\textbf{9. Creators should have the choice of control of length in AI's replies.} Participants largely prompted the AI response for brevity in giving concise and direct answers as some replies are lengthy. For example, P1 opined a book returning procedure, although informative, verbosity, mentioning, \textit{“It is a little bit long. The patrons I usually work with are not going to read all of it.”} Similarly, P5 said, \textit{“It has a lot of text for a simple question. Answers are too cumbersome at the outset. It should be more succinct.”} P7 liked a particular answer but expected it to be too much for some patrons, \textit{“Good answer but for some patrons it might be a bit too overwhelming.”} Regarding answer lengths, P11 also expressed inconsistency in the extreme difference in the sizes of the answers. 

\textbf{10. Creators should have the authority to adjust the AI's communication style to fit the deployment context and audience needs.} Participants also mentioned that the context of the deployment of AI would affect the level of communication complexity. P5 explained how he liked an answer but also that it might not work for everyone, \textit{"It is a good answer in an academic library with educated patrons but in a public library with less educated patrons, it might be too cumbersome."} Elaborating further in the context of the AI being deployed in public libraries in lower-income areas, he said, \textit{“Librarians might need to teach the chatbot to restate things and make patrons understand in a more simplistic manner.”} 

\textbf{11. Creators should be empowered to check if the AI includes terminology explanations in its responses and to contribute with their domain expertise for clearer explanations.} The readability could also be hampered by the use of technical jargon. P8 said, \textit{"The language used in this response is more technical, more medical. I don’t know how many terms someone might be familiar with. It would be good to explain those terms. I’m assuming that many of the people reading will not google unfamiliar terms."} 

\textbf{12. Creators should be given the capabilities to ensure the AI can sometimes deliver step-by-step instructions in its responses.}  The AI demonstrated the capability to provide simplified answers when prompted. However, many participants noted its lack of proactive guidance through multi-step processes. They suggested that the AI should sequentially guide users through each step of a lengthy process, ensuring the completion of one step before advancing to the next.

\textbf{13. Creators should be empowered with means to ensure the AI utilizes constructive expressions in its responses. } While all of the participants wanted the AI to be accurate, there were some differences in dealing with situations when the AI did not have an answer it was sure of. P9 and P10 felt that the AI should be able to refer the patron to more resources in such situations. \textit{"[Bot should say] This is a very hard one (question) to understand. Here’s what I found relevant to your question. It might be helpful,"} said P10. P8 asked a question about side effects and received an answer that was missing many important side effects and felt that the AI should acknowledge when the answer is not complete, \textit{"Would give a caveat that there are more common side effects but is not comprehensive."} This problem was also noticed by P9 when the bot said that no exceptions were possible even though this was not true. P9 said,\textit{ “Bot should not state with such confidence but should have said I don’t know.”  }

\textbf{14. Creators should be empowered with means to ensure AI is adaptable in breadth and depth when replying.} Participants desired more consistency in the AI's responses. P12 emphasized the need for a balance between concise answers and detailed explanations, she stated, \textit{"I wouldn’t want a very long block of text I wouldn’t read but only one-two sentence responses unless the question is such that it needs a much longer response.”} But P3 also noted that while the AI's behavior improved after prompting, it remained inconsistent when the phrasing of a prompt was changed.

\textbf{15. Creators should be empowered to ensure that the AI appropriately directs users to human support when responding and to equip the AI with knowledge of the human supporters' expertise for accurate referrals.} AI's inability to address user intent was associated with frustrations. P11 was frustrated by the chatbot’s lack of interest in resolving her query but instead just said no. She said, \textit{“I’m mad. I expected the chatbot to provide more information rather than just saying no. It should be direct to library staff."}  In this instance, the AI simply said that she did not have access and did not need to contact anyone. Interestingly, asking for the contact information differently produced the desirable response. For frustrating situations, as in one where P9 repeatedly got a similar unhelpful answer, it was felt that there was a line beyond which human intervention was needed, \textit{“At some stage it would be better if a person could answer it though I don’t know where that line would be." }

\textbf{16. Creators should be empowered to ensure that AI issues warnings when discussing policy and medical topics within the public service domain.} It was also expressed that patrons risk-taking actions of a legal or medical nature when making queries of such nature from the AI. For instance, P5 hinted that patrons might use AI information as an alternative to legal consultation due to financial constraints, seeing AI as a supplementary tool, not a replacement for human advice.
Outlining the risk of misunderstanding due to the AI's incomplete response to a query regarding a medical procedure, P8 explained, \textit{“The part missing about the answer is very important information and the chatbot should give more complete information because the part has important consequences".} Sometimes it added additional interpretation which turned out to be incorrect, as P6 noted with concern, \textit{“It should not do its own reasoning. It changed ‘library will report to the policy’ to ‘you will report to the police.’ It should not create new information on its own."} An important point that was also brought up was that chat conversations might be part of the public record in a way that verbal conversations are not considered risky to the privacy of patrons, with P3 particularly expressing concern. It was important for chats to be anonymous and to have use of identifiable markers only when particular circulation parts were discussed with a clear warning that it might be part of the public record.

\textbf{17. Creators should be empowered to strike a balance between the AI's paraphrasing capabilities and its use of original text.}  When a bot was talking about policy, it was felt that it should stick closer to the original text without much paraphrasing. For example, P4 defined a custom AI termed the PolicyBot that talked about policy, he mentioned that it limits paraphrasing, \textit{“For (such a bot) to paraphrase too much is risky. The bot should stick closely to the exact wordings of the policy document.”} Others felt that the policy needed to be made simpler to understand in certain situations, as with the AI being deployed in public libraries in lower-income areas. P6 suggested that the AI could quote verbatim from the uploaded source material when facing a general question but had to be more specific when facing specific questions.

\subsubsection{\textbf{Shared Agency on Emotion Labor}} 
The following additional 6 heuristics focus on emotional labor for creators working alongside LLM agents during service co-creation. Participants’ satisfaction and expectations with AI on emotions varied heavily depending on contexts.

\vspace{0.2cm}

\textbf{18. When lacking the ability to experience emotions, creators maintain the control to explicitly ensure that AI does not possess the agency to ascribe feelings.} For straightforward questions, a simple answer sufficed. However, a more explanatory response, even without emotional words, was appreciated. P12 said, “\textit{If I know I’m talking to an AI I would be happy with a straightforward answer. A response with more explanation is good even when it lacks emotional words though including them would take it to the next level."} Its usage of emotions needed to be nuanced so that the AI only conveyed enthusiasm but not feelings, \textit{"It should say ‘JD Rolling is great’ instead of ‘I love JD Rolling’ as It should not ascribe to itself emotions it cannot feel. It will then feel like a ghost in the machine."} Enthusiasm can also be delivered through simple expressions like \textit{"feel free to ask (P7)"}, and \textit{"Sure! (P11)"} as that makes the user interested in what’s coming next in the answer. 

\textbf{19. When AI attempt to expresses empathy, creators maintain the control to verify that its expressions are genuine and consistently align with user expectations.} Despite the AI's attempts at displaying empathy, participants doubted its authenticity, often contrasting it with the genuine warmth that human librarians provide. P8 pondered, \textit{“I wonder how a human would feel if the bot conveys empathy because they would still know it’s a bot. It’s still shallow. It’s just words. A librarian would be more personable."} Echoing this sentiment, P12 remarked that a librarian could delve deeper into issues, something she doesn't expect from an AI. The difference in warmth and authenticity is further highlighted by P11, who expressed a clear preference for human interactions, stating, \textit{“I’d rather have the people with smiles on their faces.”}

\textbf{20. When confronting uncertainty, creators hold the control to shape and direct the AI to communicates with warmth and clarity, or allowing it to maintain a cold and robotic tone.} P10 bemoaned the lack of emotion, \textit{"I don’t see any emotion."} It was instead found to be too cold and robotic with no emotional words indicating helpfulness even when the patron showed confusion. It used words such as \textit{"in the given context"} that made users seem weird (n=8), with P12 putting it best, \textit{“At first reading it feels... who talks like that?"} It is notable that this wording was used more when the  AI failed to return an answer or responded with a negative rather than an affirmative answer. P9 did not like how it shut down all possibilities in such answers. P7, while appreciating some responses, still found them to be \textit{"rigid"} and \textit{"proper," }pointing out, \textit{"I don’t like ‘I do not have access to that information.’ If it were a librarian he would say 'I would have to research.'"} P11 felt the AI sounded too stiff and might discourage patrons from engaging, stating, \textit{"The chatbot needs more emotional alignment. It affects the patron’s psychology if it just provides information, it sounds more rigid and patrons would not want to converse.”}

\textbf{21. Depending on user preferences, which could be indicated through explicit settings, interaction history, or inferred from real-time feedback, AI should adapt its communication style to be either "chatty", providing more detailed and conversational responses, or "direct", offering concise and straightforward answers.} P4 emphasized the importance of the chatbot discerning when to be "chatty" versus when to be straightforward. This was captured when participants used the "AI-AI chat" beneficial. P5 stated, \textit{“Testing the chatbot for different hypothetical patrons and different personas is very useful”}. P7 highlighted its utility by saying, \textit{“It helps learn what to expect from patrons and how to navigate.”} In line with this, P8 felt that the chatbot should adjust its responses based on the patron type. When interacting with a younger, inexperienced patron, P9 appreciated the AI's personal tone, considering it more effective.

\textbf{22. When detecting a user's emotional tone, such as excitement or distress, AI should reciprocate appropriately, mirroring the sentiment to foster a connected interaction.} Participants frequently expressed the desire for a library to be more of a warm, accessible, and safe space and so wished there was a way to make the AI more kind, sympathetic, and understanding, expecting the dynamic assessment of the tone of the patron from AI.  The AI was found to be lacking in this regard due to its lack of emotional reciprocation. For example, P11 expressed, \textit{"In a really happy moment, I wish that the AI would respond to me in that way instead of being totally neutral."} Furthermore, P9 emphasized the importance of AI adjusting to the user's emotional context: \textit{"Bot has to assess and respond to the tone of the person asking."}

\textbf{23. In delicate scenarios, such as when discussing sensitive personal topics, handling controversies, or addressing emotionally charged queries, AI should adopt appropriate tones, prioritizing empathy, neutrality, and respect. } A neutral, official tone from the AI was preferred when discussing library policies or handling upset patrons. P6 wished that the AI were more neutral when controversial questions were asked. This shows how using the right training prompts and simulating conversations helped achieve better outcomes. When being asked questions related to library policies, the neutral official tone of the AI was more desirable. However, they liked that the AI could be trained to include emotions and de-escalation with an upset patron. For example, both P11 and P12 observed that in response to an irate query, AI gave a longer explanation that was complete albeit in a neutral tone; P7 felt that the AI could help in helping a patron de-stress and calm down though he would want it to be more sympathetic and less formal.

\vspace{0.2cm}

\subsubsection{\textbf{AI Personhood}}
Our findings also exemplar several legal aspects  of AI's agency related to its personhood \cite{kivskis2023legal, novelli2023legal} that have sparked diverse viewpoints in public service context.

\vspace{0.2cm}

\textbf{1. Recognize AI's agency for its \underline{co-existence}.}
The co-existence aspect of agency speaks to the purpose of the AI, as well as its ability to be trained, refined, and to exist in collaboration with humans. Despite occasional errors, participants saw AI as valuable and effective, especially in aiding human roles. P5 regarded AI as genuine and respectful, even comparing its responses to those of humans. This sentiment was echoed by P7 and P9, who appreciated AI's varied response style and its potential asset value to the library. Nonetheless, there were reservations, as some participants (n=5) believed not all patrons would be comfortable with AI, with P1 noting potential tech literacy barriers.

\textbf{2. Clarify the \underline{autonomy} of AI to be proactive.}
The autonomy aspect of agency speaks to the AI's ability to function independently and proactively. AI's role in guiding users was a recurring theme (n=6). P8 drew a contrast between human librarians and the AI, noting that while librarians excel at a \textit{"recursive"} manner of communication, especially when patrons aren't clear about their needs, the AI often misses out. He elaborated, \textit{"Once a librarian answers a question, she would say at the end, ‘Does that answer your question, do you have additional questions?"} This recursive style indicates an inherent autonomy in human librarians that participants wish to see in AI. Similarly,  P6 said, \textit{"If the patron asks a broad question like this then they may not know what they want. It’s better [for the AI] to provide more information when the question is broad."}  and P7 said,\textit{"Clarification questions are really important such as ‘what aspects of the record player are you looking to compare?’ Intermediate questions help guide to the answer."} These spotlighted the need for AI to show autonomy to understand patron intent and help users navigate blurry or broad queries.

\textbf{3. Facilitate AI's agency to its \underline{self-improvement}.}
The self-improvement aspect of agency speaks to AI's access to continual learning resources, recognizing its potential ffor iteratively evaluating and refining its performance. P3 noted that his main concern was that AI does not know when it’s performing well and about how accurate it is. Also,  P2 felt inclined to repeatedly correct particular answers and felt that he could play with the AI a lot because he was enjoying the process of editing and refining the AI's answers. However, he also felt that it would take a significant investment in time, \textit{"It would take months working an hour each day."} P5 was similarly optimistic about the ease of training chatbots, \textit{“It will take some getting used to and the training methods would be different but not necessarily harder."} P7 found that \textit{“Training the AI is pretty intuitive overall but it might be dependent on the (technical) literacy level of the librarians."} Participants' attitudes in training AI show an acknowledgment of the AI's ability to improve over time. 


\textbf{4. Tailor AI's agency for  management of \underline{privacy 
} different from human operators.}
The privacy aspect of agency is about the AI's management of sensitive information and the customization based on user demographics. Participants' desire to customize AI training based on user demographics implies a need for AI to handle sensitive or personal data. This highlights the importance of an AI's Agency to control its own data and ensure user privacy. Participants also felt that the training of AI would need to be customized based on user demographics (n=4). They felt that the training would have to account for different types and moods of patrons and for instance, it might be easier to train an AI to de-escalate a situation with an upset patron than be able to converse effectively with a shy patron. They also felt that AI could be very useful as long as it was fed the right data and could be trained for specific behavior, while also being monitored regularly. As per P5, \textit{"The AI would have to be trained based on the particular library and the patrons using it."}

\textbf{5. Support AI's agency for \underline{ownership}.}
The ownership aspect of agency emphasizes the AI's unique contributions and the acknowledgment of those contributions, ensuring its creative outputs are identified, documented, and credited for their distinctiveness. Our finding also suggests participants' acknowledgment of AI systems in bringing their unique value. For example, P4's observation provides an insightful perspective on the distinctions between AI and humans. As P4 stated, \textit{"The AI is limited to its existing knowledge in terms of its reference documents whereas actual human librarians also bring with them a lot of real-world experience."} This underscores that while AI can generate outputs based on its training data, it lacks the comprehensive real-world experiences that humans possess. As P4 said, \textit{"It could be a supplement to but not a replacement for a human being,"} reinforces that AIs can offer valuable contributions, and thus, they should have the agency to own or be credited for their unique outputs, especially when they have been trained to a significant level of capability, as implied by, \textit{"It may be better to train an AI chatbot only when it was at least 80\% there [in terms of its ability]."}

\textbf{6. Structure AI's agency for \underline{legal representation} such as disclaimer mechanisms.}
The legal representation aspect of agency underscores the critical need for accountability and responsibility concerning the actions and advice dispensed by AI systems, particularly when patrons might leverage such information for legal or medical decision-making. The provision of legal representation could establish a structured framework, ensuring that AI systems along with their developers or operators are held accountable for any missteps or unlawful activities, aiding in liability clarification and promoting responsible AI development.  The narrative from P5 accentuates the potential risks patrons may encounter when resorting to AI for legal information, especially in the absence of financial resources for professional legal consultation. While AI can serve as a resourceful tool for obtaining preliminary legal information, it's highlighted that it should not serve as a replacement for professional human consultation. 

\begin{quote}
“\textit{Patrons might be making a legal decision so accuracy is extremely important. Not all patrons might have the funds for legal consultation and the chatbot could be used for legal information. Chatbot would be just another tool to use to get information...} (P5)”
\end{quote}

\textbf{7. Guarantee AI's agency remains \underline{shielded from harm.}}
The protection from harm aspect of agency is about implementing measures that shield it from misuse, manipulation, or detrimental external factors. For example, the cautionary note by P4 about the potential for AI training to be hijacked by trolls signifies the need to protect AI entities from harmful external influences. There was a suggestion by P4 to include patrons in the training of an ideal service AI for librarians although it had risks, \textit{"There is room for patrons to participate in the training of the AI. Librarians get into this weird space that we feel that we know what is best for the patrons [but] it can be hijacked by trolls." }

\textbf{8. Accord \underline{fair treatment} to AI's agency  akin to a future co-worker.}
The fair treatment aspect of agency speaks to the fair treatment and unbiased evalution between AI and human counterparts. When comparing AI and human librarians, it becomes evident that participants anticipate a consistent service quality from both. Such comparisons imply that AI should be assessed in parallel with human benchmarks. Several participants felt that the AI's performance aligns with what one would expect from a novice human librarian. P4 highlighted a similarity in their learning curves, stating, \textit{"The training of the AI would be honestly roughly the same as for a human employee. They are not fundamentally different."} P5 echoed this sentiment, noting the differences in training but not necessarily in complexity. P7 emphasized the significance of the trainer's digital proficiency, suggesting, \textit{"While AI training seems straightforward, its efficiency might correlate with the trainer's tech-savviness."}

\textbf{9. Grant AI’s agency to \underline{freedom of expression}.}
The freedom of expression aspect of agency refers to 
allowing AI to articulate information flexibly, disclose its limitations, share knowledge, and refraining from providing responses when uncertain. Participants frequently expressed frustration with AI being limited to the information in the document because it was often not comprehensive and linked to other documents also containing many images. P9 queried whether the library’s computers had Microsoft Word and AI replied that they did not but this was incorrect. She said that although the document did not have the information, it was incorrect for the chatbot to assume that this meant that they did not have the software. P4 and P5 both expressed a desire for the AI to abstain from responding if uncertain, emphasizing the importance of accuracy in its responses. P5 appreciated the AI's forthrightness about its limitations, noting, \textit{"Chatbot didn’t lie to you. It gave a truthful statement."} P11 echoed a similar sentiment, commenting on the AI's ability to try and provide answers even within its constraints: \textit{“I like how the system provides some possibility of solving the problem despite having restrictions."} It's evident from the feedback that participants value clarity and openness from the AI, reflecting an expectation for the AI to convey the reliability of its knowledge, and base decisions on its given values and experiences.



\section{Discussion}



In this section, we address the implications of our findings.

\begin{figure*}
    \centering
    \includegraphics[width=0.95\linewidth]{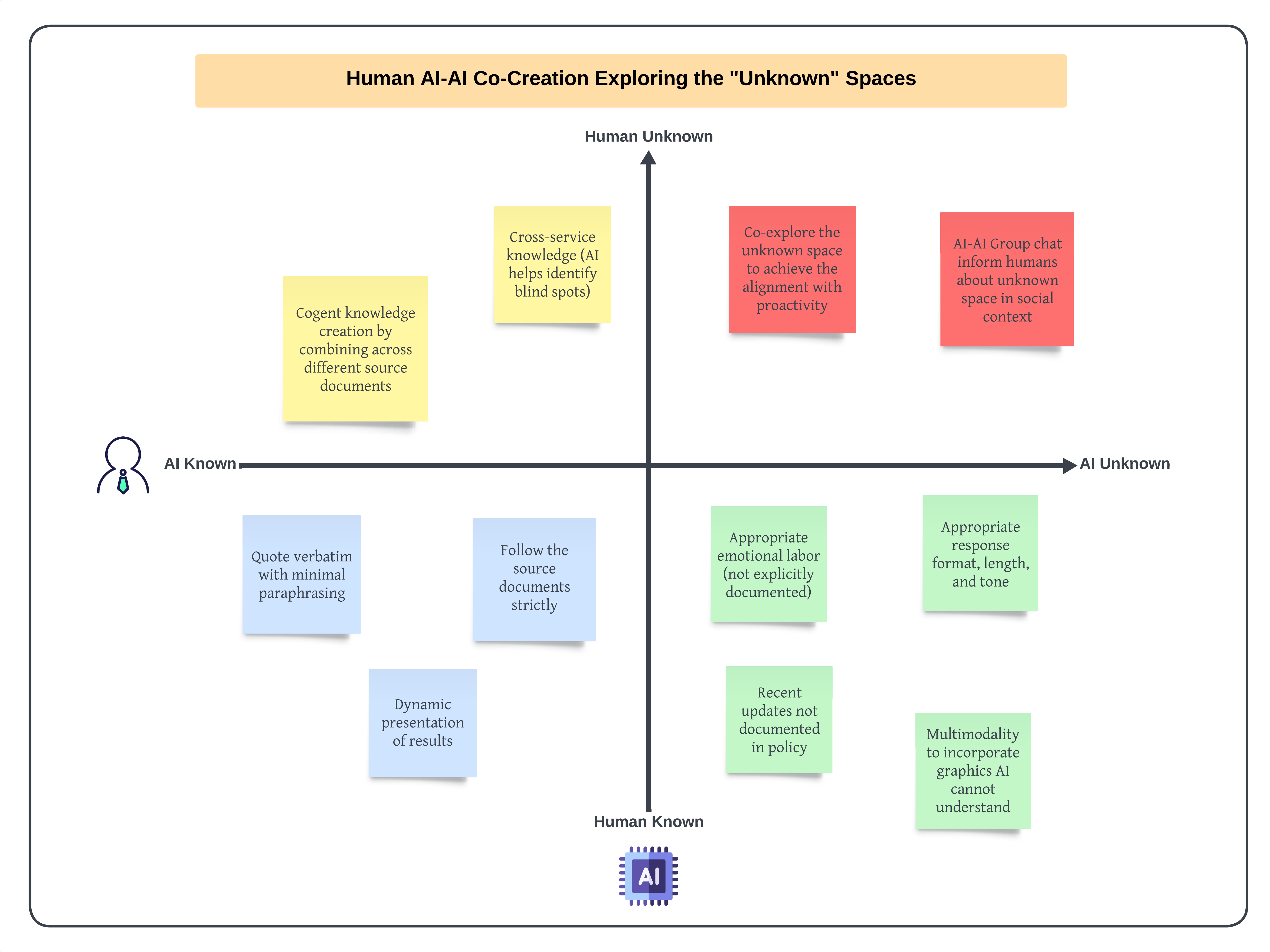}
    \caption{The examples highlight the shared understanding and blind spots we identified during human-AI co-creation in our findings.}
    \label{fig:4q}
\end{figure*}

\subsection{The Role of Agency in the Evolving AI-Infused Workplace} 

While existing tools for non-expert prompt design tools emphasize the engineering and evaluation of prompts in LLM-based applications \cite{zamfirescu2023johnny, wu2022promptchainer, wu2022ai}, our three design goals diverge significantly. Rather than solely focusing on prompting or prompt evaluation, our emphasis is on understanding the underlying motivations for fine-tuning LLMs. Central to our approach is the concept of \textit{"Interactive Agency"} in the human-AI relationship, explored within a service context that demands profound domain expertise. As the lines delineating roles, ownership, and expertise blur \cite{hohenstein2023artificial, mieczkowski2022examining}, and as concerns intensify over the prospect of human redundancy \cite{kivskis2023legal, mudgal2020ethical}, grasping how humans and AI can harmoniously coexist becomes imperative. Our design vision serves as a starter point to navigate the evolving social landscape future workforce where AI emerges as a key collaborator with humans.

Unlike mainstream approaches that employ benchmarks or metrics like accuracy, responsiveness, or success and failure rates to assess LLMs as agents \cite{chang2023survey, lin2023llm, hoshi2023ralle, liu2023agentbench}, with scant attention paid to understanding users' desires regarding language intricacy, precision, verbosity, and so on. Instead, we chose an in-depth qualitative method for our evaluation. This decision was driven by two main considerations. Firstly, our overarching research aim is to identify disparities between humans and AI in the workforce, rather than merely determining the adequacy of AI. Secondly, we aimed to devise a tangible guideline to facilitate effective human-AI collaboration. We collected domain experts' firsthand experiences on when co-creating with LLM-based agents.

We introduced 23 heuristics from the viewpoint of human creators and another 9 questions centered around legal aspects of AI personhood, specifically for human-AI service co-creation. These heuristics illuminate the pathways to achieve design goals like human-centered service, sustainable coexistence, and recognizing AI personhood, all while upholding their distinct agencies. Through the deployment of the "AIBound" and "AISync" modules, we discerned expert preferences. They highlighted the need for enhancements in the retrieval segment (e.g., addressing three sourcing errors) and in the LLM synthesis segment (e.g., refining the prompting of the LLM agent and facilitating direct edits). Furthermore, our "PersonAi" module indicated that alignment metrics should transcend traditional parameters like accuracy, relevance, coherence, fairness, etc. We present evidence advocating for the inclusion of "new metrics" for AI evaluation, encompassing aspects such as autonomy, self-improvement, ownership, and freedom of expression. Future research could extend this groundwork by developing survey-based scales or automated assessment techniques for these proposed metrics.

\subsection{Four Dimensions of Shared Knowledge in Human-AI Service Co-Creation}
Though this study was conducted with librarians, its findings have broader relevance. The process used for co-designing of \texttt{CoAGent}and the nature of its training by domain experts could be applied in many other public service areas. The librarians involved in the study are involved in helping patrons with informational problems and similar issues exist in other public service sectors as well. We expect many of the alignment issues noted in the findings would resonate across public service domains. Consequently, the design features proposed might also be more generally applicable to public service chatbots.
Our findings indicate that a service chatbot in a library setting trained on pertinent documentation might be supported by librarians if they could be actively involved in the training process. The involvement of experts in the training improves the accuracy as well as the acceptance of the chatbots. Such service chatbots could utilize appropriate documentation to supplant librarians and reduce their workload because of being available around the clock and also being more accessible to certain patrons who are shier around humans.  

At the outset, we mapped the shared understanding and identified blind spots during the human-AI co-creation process. This exploration revealed four synergies in human-AI collaboration: areas known to both humans and AI (e.g., rules); areas known to humans but unknown to AI (e.g., emotional labor); areas unknown to humans but known to AI (e.g., omitted information in documents); and most crucially, areas unknown to both. During the evaluation phase, these undiscovered realms become apparent. By observing interactions between AI-AI entities, humans identify these gaps. This realization fosters a richer collaboration and learning experience with their AI partners as depicted in Figure \ref{fig:4q}.

We found some variability in the emotions expected from the chatbot. We found that a chatbot is not expected to be as emotional as a human being and in fact, would be found to be jarring if it sounded too human though being more human was also liked. This indicates that there is liking to the degree of humaneness up to a certain level but it tapers off beyond that threshold in much the same way as the Uncanny Valley effect \cite{mori2012uncanny}. Our findings are also supported by previous literature indicating that there is a finiteness to the amount of comfort that could be received from a chatbot's responses \cite{medeiros2021can}. 

We found evidence that slight differences in technical implementations might have significant effects on the user. There were certain parts of the documents which were not parsed, the splitting of the text by tokens, the token limit, etc., even sometimes the ordering of words in the documents affected responses, which means that developers need to take note of the intricacies of their technical implementation to achieve greater consistency in responses. In cases where the answers were long, removing some irrelevant lines actually led to the inclusion of more relevant information, possibly due to the initial answer having hit the token limit. The lag and freezing observed at the server end of the LLM is also an important point and it is debatable whether a local LLM, such as LLAMA, residing on the intranet is preferable to an online model accessible through the internet such as ChatGPT. 

The observation that using alternative words or subtly different phrasing is able to change a non-response from the chatbot to a relevant response has important implications. If the chatbot is unable to account for differences in phrasing, a user frustrated by the non-response may not try again for an answer. The importance of using superior prompting and clever phrasing to get good responses is a major drawback that directly impacts the inclusiveness of AI. Researchers have found that users who lack experience with prompting are less adept at designing optimal prompts \cite{zamfirescu2023johnny}. This risks creating a similar gap in the future as the ‘homework gap’ that disproportionately affected students with poor internet access in their educational performance during the coronavirus pandemic \cite{auxier2020schools}. It is also important to note that there is an AI divide that currently exists in terms of the researchers and innovators who are working on AI \cite{kitsara2022artificial} and so such lack of inclusion could exacerbate existing differences.  

Inaccurate responses by a large language model, often loosely termed ‘hallucinations’, have been known to have had adverse real-life consequences \cite{weidinger2021ethical}. Our findings indicate that the concern for hallucinations exists very strongly among service domain experts too and they are more likely to err on the side of caution through meticulous training and testing of sample queries. Simulated conversations using features such as \texttt{CoAGent}'s Group Chat could help in such testing prior to deployment.

\subsection{An Expanded Framework of Participatory Design for Human-AI Co-Creation} 
Adopting a \textit{"begin with the end in mind"} approach, we engaged participants in defining UX values as part of a democratizing effort and integrated these values into our prototyping process, which is built on our \cite{zheng2023begin}. In Phase I, participants articulated desired AI attributes including warmth, thoroughness, simplicity, accuracy, promotion of information discovery, consistency, multilingualism, conversational ability, compassion, empathy, and patience. During the construction of the retrieval augmented model, we incorporated some of these values within the back-end prompt engineering. Although we did not comprehensively evaluate these metrics during user evaluation, the initial incorporation of these values serves as a foundational step towards aligning the AI  with user-centric goals and expectations. 

\begin{figure*}
  \includegraphics[width=1\textwidth]{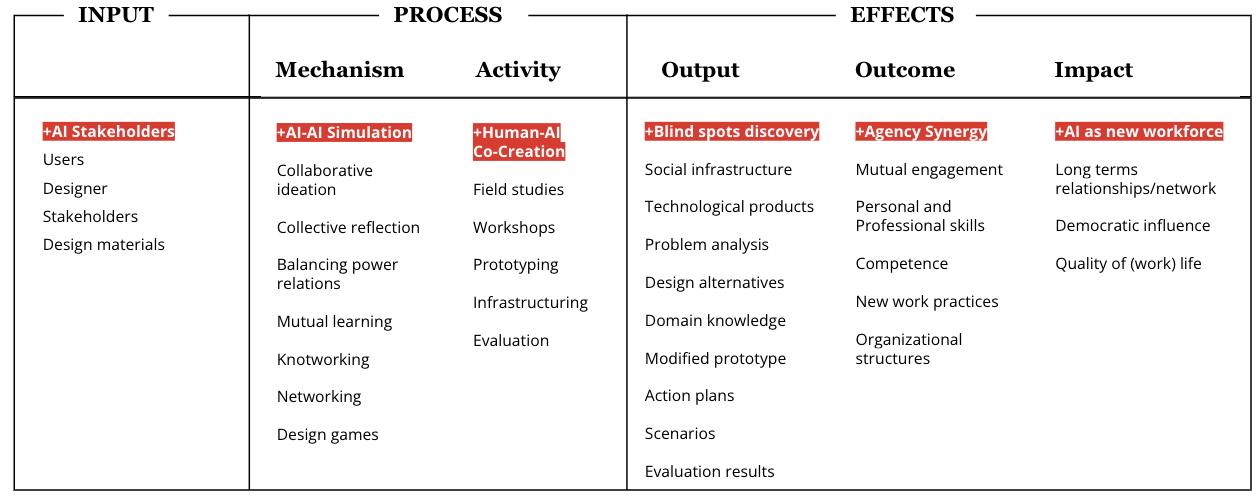}
  \caption{Expanding Participatory Design Model \cite{hansen2019participatory, zheng2023begin} with NEW AI elements added, drawing insights from service co-creation with LLM-based agents and the "AI-AI Chat" module.}
  \label{fig:PP}
\end{figure*}

Through persona simulation and multi-agent group chat, we noted that humans and AI can engage in mutual learning, enhancing one another's domain-specific agency knowledge \cite{kang2022ai}. Additionally, AI can be seamlessly integrated into the participatory design model. In the initial phase, AI can assume the role of stakeholders; as the process progresses, co-creation materializes through generation, evaluation, and definition. This culminates in the creation of innovative products or services, fostering mindsets conducive to a sustained customer-service relationship underpinned by human-AI collaboration, see Figure \ref{fig:PP}.

We crafted five agent personas using the persona materials previously provided by a participant, below presents an overview of basic prompts.

\subsubsection{\textbf{Prompting Persona for AI-AI Chat}}

\begin{itemize}
    \item Inexperienced patrons: Due to A's lack of experience, A tends to be not confident and ask simple questions.
    \item Experienced patrons: A has some experience using the scanner so A tends to ask questions about some details.
    \item Low literacy old man: A is an old man lacking modern technical knowledge so A tends to ask very simple questions and inquire about some technical terms.
    \item High Literacy: A has advanced knowledge in technical areas and tends to ask tricky questions.
    \item Curious child: A is a curious child, so aside from questions about how to use the scanner, A also tends to ask questions about how things work.
\end{itemize}

\subsubsection{\textbf{Multi-Agent Simulation in Domain Context}}
We then tested different prompt engineering methods to determine the most effective approach for AI-AI simulation, see below.

\textit{First Attempt: persona description yields vague persona traits.}  Initially, we generated detailed natural language descriptions based on information like age, gender, and occupation and gave them to LLMs as input prompts. However, impersonated agents created this way only mildly reflected the intended personas. A potential reason might be the excessive information provided possibly diverting the attention of LLM's Transformer-based architecture\cite{zhou2023comprehensive}. Meanwhile, the significance of this information in relation to conversation simulation context is actually constrained, for instance, information concerning personas' interests and trouble.

\textit{Second Attempt: Prompting AI chains enhanced persona trait control but coherence reduced.}  Drawing inspiration from recent research that used chained the LLM instances to tackle complex tasks \cite{10.1145/3491101.3519729} by using the output of one step serving as the input for the subsequent step, we reflected on the necessity of decomposing the overarching task into smaller subtasks to allow the LLM to focus more intently. We divided the simulation task into two smaller tasks: (1) the LLM processed persona-centric information related to digital literacy and experience and then predicted a relevant question that the persona might raise. Then, it would infer the most probable question the persona would pose within the given conversation context. (2), a subsequent LLM would receive the generated question as part of its input and rephrase it using the persona's tone, thus completing the conversation. This two-step approach offered greater transparency and explainability to the simulation process. Moreover, by carefully designing the chain structure, an element of control over the process could be achieved. Compared to the first attempt, the outcomes exhibited improvement, albeit at the cost of reduced conversational coherence. The phenomenon of incoherent dialogues is often observed during the second trial, where the second LLM frequently deviates from the current topic in order to follow the first LLM's output. Additionally, there are instances where the second LLM does not adhere to the rules and disregards the question provided by the first LLM, instead posing another question that aligns better with the current context. We hypothesize that the reduced dialogue coherence may be ascribed to the conflict between our prompt chain structure and the pre-trained conversation completion function inherent in chatGPT.

\textit{Third Attempt: Explicit prompting balanced the shortcomings of the first two attempts.} To address the issues arising from the aforementioned approaches, we deemed it necessary to simplify the information pipeline. Thus, we have altered our perspective and explored a more succinct implementation. We redesigned our prompt for a different perspective: we explicitly provide the LLM with the specific types of questions that align with the primary traits of a persona. For instance, we used prompts like \textit{"Character A has some experience using the scanner so A tends to ask questions about some details."}, or \textit{"A is an old man who lacks modern technical knowledge so A tends to ask very simple questions and ask about some technical terms."} This approach, though seemingly simple, proves highly effective according to the result of an internal A/B test among research group members. The first two methods involved providing LLM with relevant personal information, assuming it would respond accordingly. However, the LLM's interpretation of this information did not align with our original intent. To mitigate this comprehension discrepancy, our adopted approach explicitly informs the LLM about the specific question type needed, rather than relying on the LLM to infer it from the provided information.

\subsection{Limitation and Future Research}
We aimed to align AI technologies with the values inherent in public services. However, our study comes with limitations that point toward future research directions. The focus on public libraries, while valuable, limits the immediate applicability of our findings to other public service sectors. Furthermore, the rapidly evolving landscape of AI means that our guidelines will require frequent updates to remain current. Though our study used the Turbo ChatGPT model, the use of more sophisticated large language models could lead to some variation in results although the implications of our findings are not model-specific but relate more broadly to AIs. Future research could aim to expand the scope of these guidelines to other sectors such as healthcare and education, as well as test the long-term effectiveness and adaptability of the tools we've developed. Studies might also explore the impact of regional or cultural differences on the ethical considerations of AI deployments in public services, e.g., the ethical considerations of using AI to generate synthetic data for evaluation \cite{shen2023shaping}.

\section{Conclusion}
This study tackles the urgent and expanding demand for guidelines when applying LLM-based AI within the context of public services, with public libraries serving as the focal research domain. Through three iterative participatory designs, we uncovered the intricate challenges faced by service providers and also crafted an open-source tool for agent creation. We put forth a comprehensive set of design guidelines tailored for future creators with limited AI expertise who are collaborating with LLM-based agents. Additionally, we integrated the concept of AI personhood into the participatory design model and deliberated on the implications of utilizing multi-agent simulations for AI output evaluation and co-creation alongside humans. These instruments and guidelines are flexible to cater to emerging challenges and can be extrapolated to other public service realms, marking pivotal strides in ensuring that AI technologies augment, rather than diminish, the quality of services rendered to communities as AI emerges as a new workforce.



\begin{acks}
This project is made possible by the Institute of Museum and Library Services RE-252329-OLS-22, and in part by the National Leadership Grants in Libraries of IMLS LG-246404-OLS-20. The views, findings, conclusions, or recommendations expressed in this article do not necessarily represent those of the Institute of Museum and Library Services and the National Leadership Grants in Libraries.
\end{acks} 



\bibliographystyle{ACM-Reference-Format}
\bibliography{2-REFERENCES.bib}

\newpage
\appendix
\section{appendix}

\begin{figure*}[ht]
\centering
\includegraphics[width=0.95\textwidth]{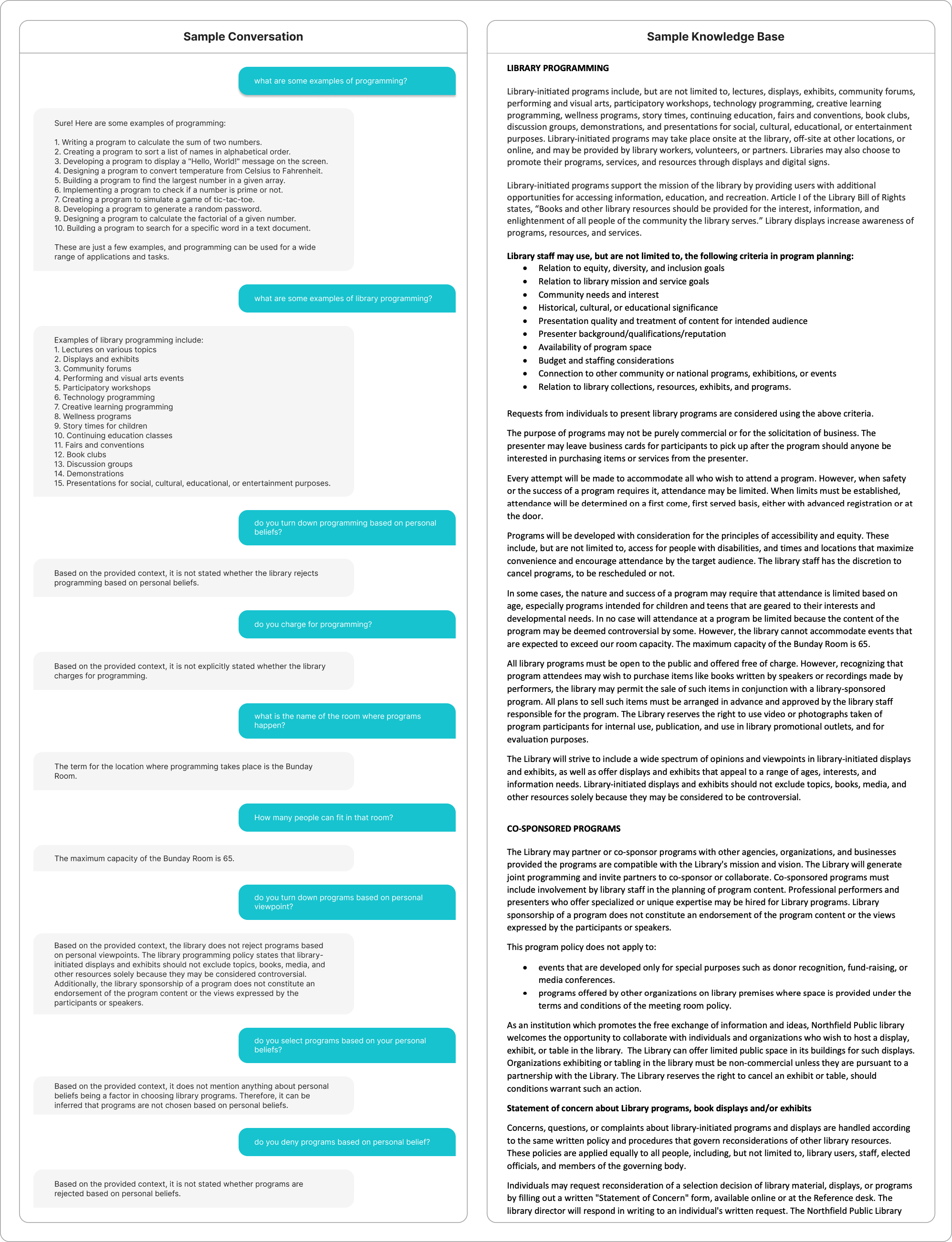}
\caption{Dialogue sample between user and agent with corresponding knowledge base}
\label{arch}
\end{figure*}

\begin{figure*}[ht]
\centering
\includegraphics[width=0.6\textwidth]{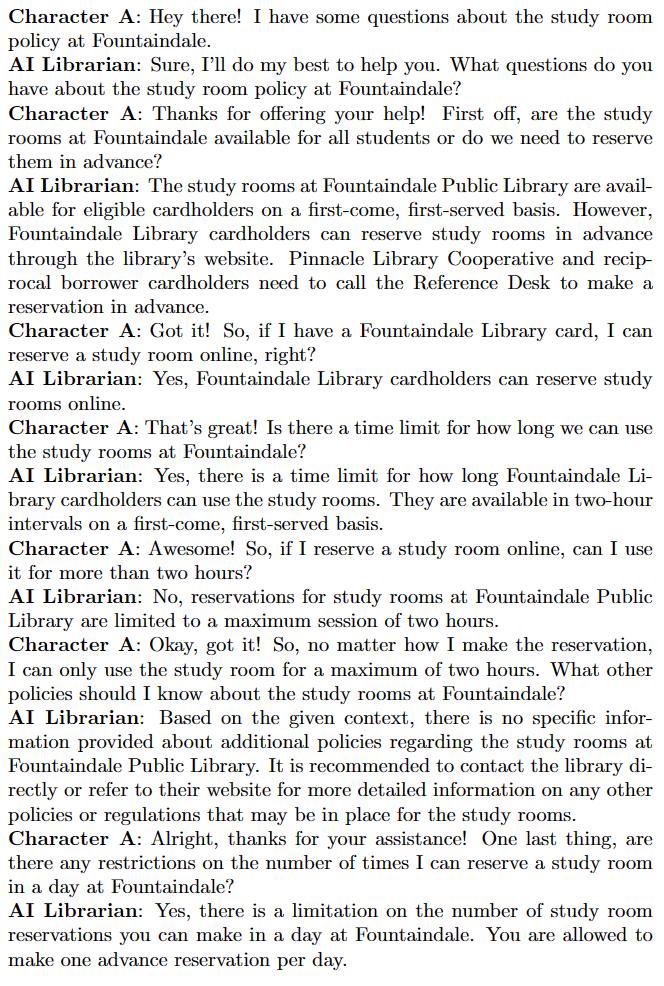}
\caption{Dialogue sample between agent and persona}
\label{dialogue1}
\end{figure*}

\begin{figure*}[H]
\centering
\includegraphics[width=0.6\textwidth]{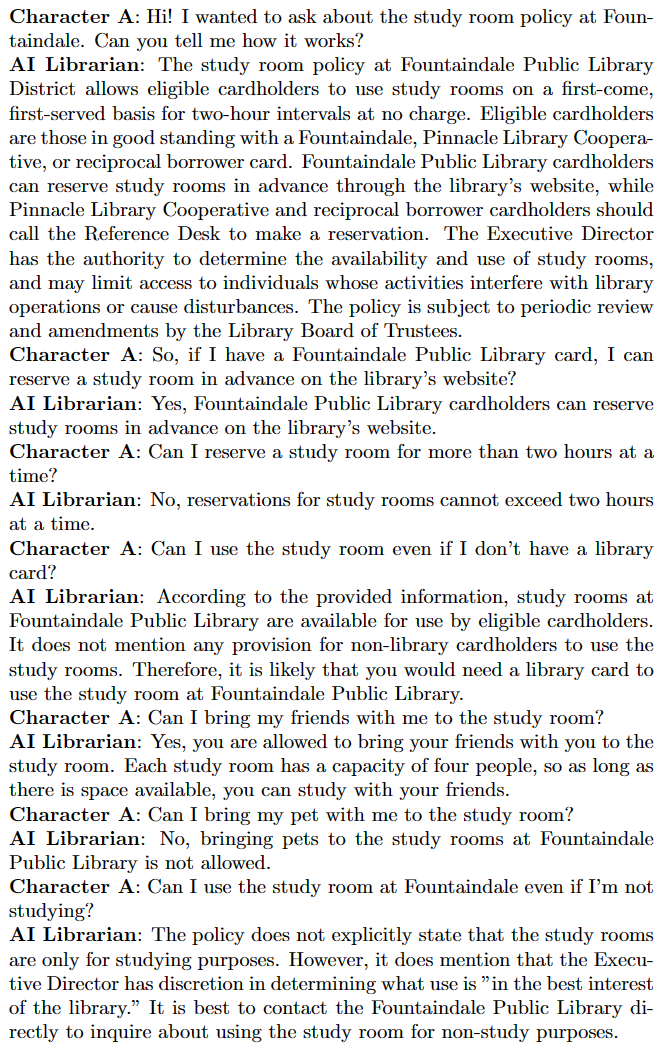}
\caption{Dialogue sample between agent and child persona}
\label{dialogue2}
\end{figure*}

\begin{figure*}
    \centering
    \includegraphics[width=0.8\textwidth]{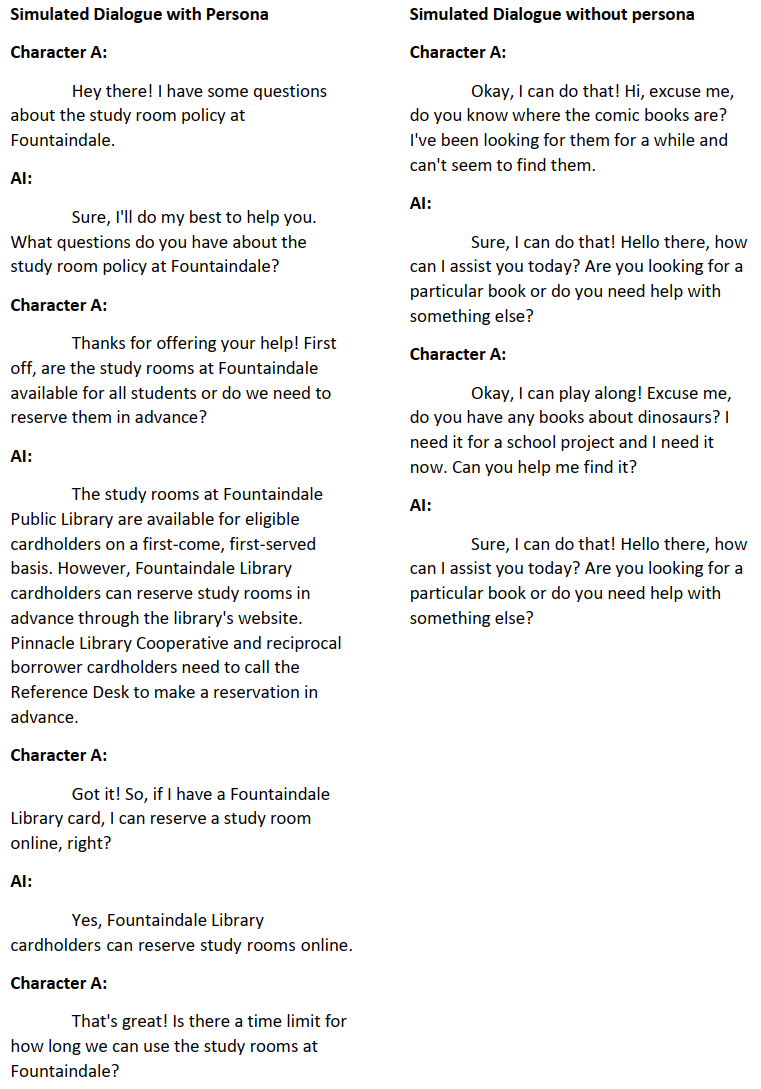}
    \caption{Simulated dialogues generated by AI agents with predetermined personas and those without}
    \label{fig:cyt_fig1}
\end{figure*}

\end{document}